\documentclass[12pt,preprint]{aastex}
\shorttitle{Silicon enrichment in planet-host stars}
\shortauthors{Robinson et al.}
\begin{document}

\title{Silicon and Nickel Enrichment in Planet-Host Stars: Observations
and Implications for the Core-Accretion Theory of Planet Formation}

\author{Sarah E. Robinson\altaffilmark{1},
Gregory Laughlin\altaffilmark{1},
Peter Bodenheimer\altaffilmark{1}, and
Debra Fischer\altaffilmark{2}}

\altaffiltext{1}{University of California Observatories/Lick
Observatory, Department of Astronomy and Astrophysics, University of
California at Santa Cruz, Interdisciplinary Sciences Building, Santa
Cruz, CA 95064; ser@ucolick.org, laughlin@ucolick.org,
peter@ucolick.org}

\altaffiltext{2}{Department of Physics \& Astronomy, San Francisco State
University, San Francisco, CA 94132; fischer@stars.sfsu.edu}


\begin{abstract}

We present evidence that stars with planets exhibit statistically
significant silicon and nickel enrichment over the general metal-rich
population.  We also present simulations which predict silicon
enhancement of planet hosts within the context of the core-accretion
hypothesis for giant planet formation.  Because silicon and oxygen are
both $\alpha$-elements, [Si/Fe] traces [O/Fe], so the silicon
enhancement in planet hosts predicts that these stars are oxygen-rich as
well.
We present new numerical simulations of planet formation by core
accretion that establish the timescale on which a Jovian planet reaches
rapid gas accretion, $t_{\rm rga}$, as a function of solid surface
density $\sigma_{\rm solid}$: $(t_{\rm rga} / 1 \; {\rm Myr}) = (
\sigma_{\rm solid} / 25.0 \; {\rm g \: cm}^{-2} )^{-1.44}$.  This
relation enables us to construct Monte Carlo simulations  that predict
the fraction of star-disk systems that form planets as a function of
[Fe/H], [Si/Fe], disk mass, outer disk radius and disk lifetime.  Our
simulations reproduce both the known planet-metallicity correlation and
the planet-silicon correlation reported in this paper.  The simulations
predict that $16\%$ of Solar-type stars form Jupiter-mass planets, in
agreement with $12\%$ predicted from extrapolation of the observed
planet frequency-semimajor axis distribution.
Although a
simple interpretation of core accretion predicts that the planet-silicon
correlation should be much stronger than the planet-nickel correlation,
we observe the same degree of silicon and nickel enhancement in planet
hosts.  If this result persists once more planets have been discovered,
it might indicate a complexity in the chemistry of planet formation
beyond the simple accumulation of solids in the core accretion theory.


 \end{abstract}


\section{Introduction}

Since the first recognition that metal-rich stars are more likely to
harbor planets (Gonzalez 1997), there have been tantalizing suggestions
that planet hosts undergo a different process of chemical development
than planetless field stars.  \cite{gonzalez97} proposed that stars with
planets are metal-rich because they self-pollute by planetesimal
accretion during the planet-formation epoch.  \cite{sandquist98} found
that accretion of a few tens of earth masses of planetesimals would
account for all the metals present in the convective envelope of
solar-type stars, but the giant-planet migration mechanism for
scattering such a substantial planetesimal mass into the host star is
only possible if $90\%$ of close encounters result in the outward
ejection of planetesimals.  \cite{mc02} confirmed that stars with
planets are iron-rich, and proposed that accretion of $5 M_{\earth}$ of
iron, in addition to high primordial metallicity, would explain the
trend.  The self-pollution scenario leads to the expectation that planet
hosts, although being iron-rich, have reduced abundances of volatiles
such as C, N and O, because these would be minimally present in the
accreted planetesimals (Smith et al. 2001).  Finding no evidence of
volatile depletion among stars with planets (Ecuvillon et al. 2004,
Takeda \& Honda 2005), most investigators have concluded that the iron
enrichment of planet hosts is primordial: stars hosting giant planets
form preferentially in metal-rich molecular clouds.  This assessment
concurs with the finding of \cite{FV05} that there is no relation
between metallicities of planet hosts and convection zone depth, either
while the star is on the main sequence or after it evolves to the
subgiant branch and its convection zone deepens.  If pollution of stars
by planetesimal accretion were responsible for the planet-metallicity
correlation, planet hosts with the most shallow convection zones would
have the highest metallicity (Laughlin \& Adams 1997).

Barring planet host self-pollution as the chief cause of the
planet-metallicity correlation, there is still the possibility that
stars with planets form from molecular clouds with anomalous metal
enrichment histories.  \cite{gchem} calculated the enrichment pattern
over time of metals from Li to Zn in sequences [X/Fe] vs. [Fe/H].  For
most elements, these theoretical sequences closely match observations.
If planet hosts do not lie on the same [X/Fe] vs. [Fe/H] sequences of
chemical evolution as other Pop I stars, the enhanced iron abundance
seen in planet hosts could be an artifact of this unusual chemical
evolution.  If the ability to form planets were uniformly the result of
a particular type of chemical event--a nearby supernova, for
example--planet hosts would show markedly different abundance
distributions than field stars in both volatile and refractory elements.
Yet analysis of the trends in [X/Fe] vs. [Fe/H] for $\alpha$- and
Fe-peak elements by \cite{bodaghee03} reveals that abundance
distributions of planet hosts are the high-metallicity extensions of the
trends governing chemical evolution in planetless stars.  A similar
analysis by \cite{huang05} suggests that, although S \& Mg may be
enhanced in planet hosts (contrary to Ecuvillon et al. 2004, who found
no evidence of sulfur enhancement), they still follow the same slope in
the [S/H] vs. [Fe/H] relation as the comparison sample of planetless
field stars.  We are left, then, with the conclusion that planet-bearing
stars are indistinguishable from other Population I stars in their
chemical enrichment histories, implying that no extraordinary chemical
events are necessary to stimulate planet formation.

Working with the assumption that planet hosts belong to the same stellar
population as other metal-rich field stars, is it still worth checking
for patterns in the abundances of stars with planets?  The studies
mentioned above use comparison samples of planetless field stars with
markedly lower median [Fe/H] than that of the planet hosts.
\cite{bodaghee03} and \cite{ecuvillon04} have no stars in their
comparison sample in the range $0.2 \leq {\rm [Fe/H]} \leq 0.5$, which
is where more than $1/3$ of their planet hosts are found.  Using a
comparison sample with no stars of ${\rm [Fe/H]} \geq 0.1$,
\cite{huang05} could not tell if the Mg and S enhancement in their
planet hosts was particular to stars with planets, or a characteristic
of metal-rich stars in general.  While comparison samples of this type
allow the investigator to comment on the general chemical evolution
trends of planet hosts, they cannot elucidate whether planet hosts
systematically possess more or less of a certain element than field
stars {\it of the same metallicity.}  Comparison between planet hosts
and field stars of the same metallicity probes a third type of
relationship between element abundances and the presence or absence of
planets: stars with planets do not self-enrich at birth, are not born in
molecular clouds with chemically tumultuous histories; but they
nevertheless may be enhanced or deficient in some element that aids the
formation of planets, {\it simply due to the natural Galactic variation
in stellar abundances.}

The chemical compositions of metal-rich stars are strikingly uniform: in
the sample of metal-rich stars observed by \cite{SPOCS}, [X/Fe] at a
given value of [Fe/H], where X = Na, Si, Ti and Ni, is observed to vary
by 0.4 dex at the very most (Valenti \& Fischer 2005).  However,
\cite{FV05}, in their study of the planet-metallicity correlation, find
that the probability of planet detection increases by a factor of five
if when iron abundance is enhanced by a factor of two (0.3 dex).  If a
correlation of this magnitude exists between planet detectability and
another element besides iron--or if iron abundance derives its power as
a predictor of planet presence from its correlation with the abundance
of another element of more physical importance to the planet-formation
process--it will be detectable in the 0.4 dex spread of [X/Fe] in stars
with the same value of [Fe/H].

In this work, we present our statistical method for assessing the
relationship between planet presence and the observed abundances of
individual refractory elements. As specific examples of the method, in
\S \ref{observed}, we present statistical evidence of possible silicon
and nickel enhancement in planet hosts, and we show additionally that
titanium abundances display no such correlation.  In \S \ref{expsim}, we
empirically determine the exponent of a power law giving probability of
planet formation as a function of [Si/Fe], among stars of the same
[Fe/H].  Finally, in \S \ref{model}, we construct a simulation based on
the core accretion theory that reproduces both the planet-metallicity
correlation and our observed correlation between silicon abundance and
presence of planets.

\section{Observational Evidence for Silicon and Nickel Enhancement of
Planet Hosts}
\label{observed}

The Spectroscopic Properties of Cool Stars catalog (Valenti \& Fischer
2005, hereafter SPOCS) is a collection of stellar parameters for 1040
nearby FGK dwarfs, including 99 planet hosts\footnote{Valenti \& Fischer
observed Vesta to obtain a Solar spectrum, so we count the Sun as a
planet host present in the SPOCS data.}, that were observed as part of
the Keck, Lick and AAT planet-search programs.  The SPOCS data are
uniform, since all observations were obtained by the same observers and
analyzed with the same spectral-synthesis pipeline.  SPOCS report
abundances of five elements: sodium, the Fe-peak elements iron and
nickel, and $\alpha$-process elements silicon and titanium, which were
measured with high precision: $\sigma_{\rm [Na/H]} = 0.032$ dex,
$\sigma_{\rm [Fe/H]} = 0.030$ dex, $\sigma_{\rm [Si/H]} = 0.019$ dex,
$\sigma_{\rm [Ti/H]} = 0.046$ dex, and $\sigma_{\rm [Ni/H]} = 0.030$
dex.  The low spread in element abundances at a given value of [Fe/H]
means precision is paramount in any successful study of the chemistry of
planet hosts.  Because of the low random errors and uniformity of the
metal abundances presented in the SPOCS catalog, it is an ideal arena in
which to study differential abundances in planet-bearing stars.

\begin{figure}
\plotone{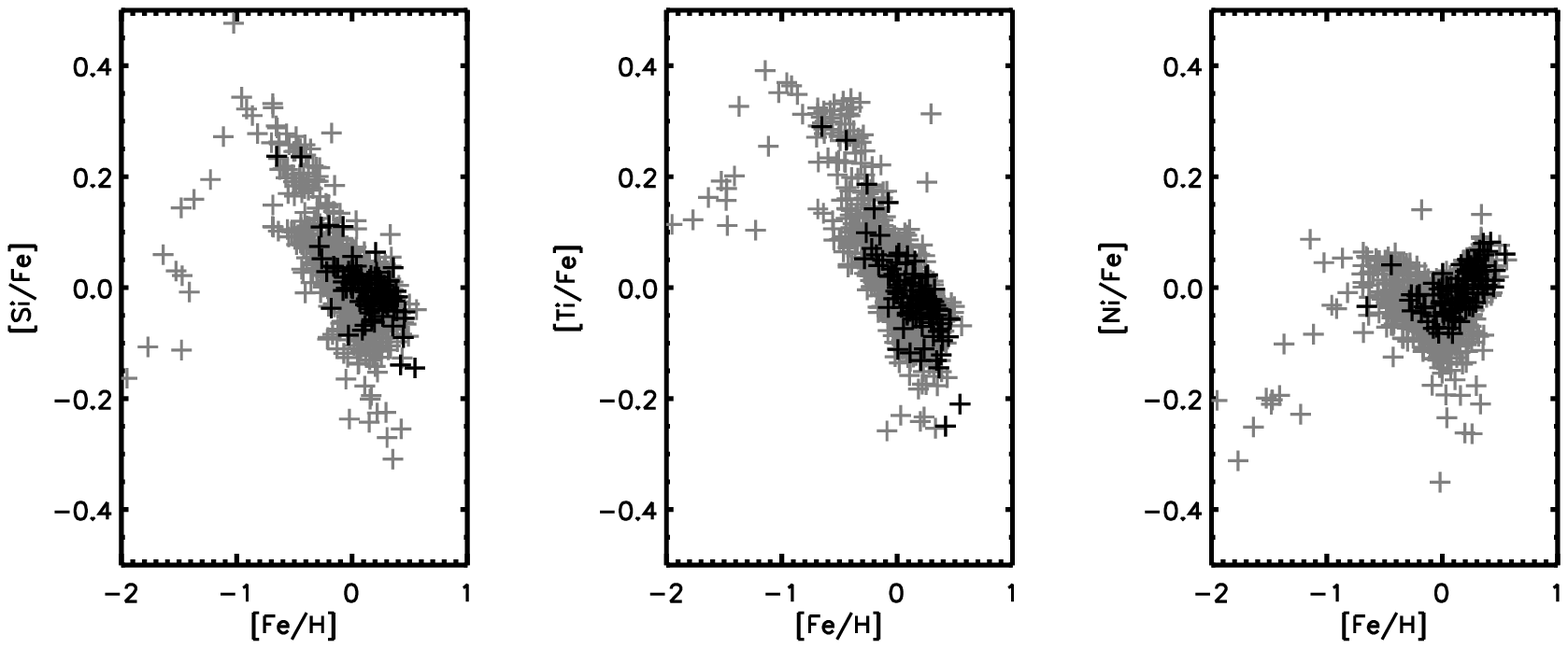}
\caption{[Si/Fe] (left), [Ti/Fe] (middle), and [Ni/Fe] (right) as a
function of [Fe/H] for the stars in the SPOCS data set.  Known planet
hosts are plotted in black and stars with no detected planetary
companion are plotted in gray.}
\label{siscatter}
\end{figure}

Figure \ref{siscatter} shows plots of [Si/Fe], [Ti/Fe] and [Ni/Fe] as a
function of [Fe/H] for the 1040 stars in the SPOCS compilation.  In
Figure \ref{siscatter}, it appears that there is a difference in the
silicon-to-iron ratio of planet hosts compared to the rest of the SPOCS
stars: Planet hosts seem to concentrate in the silicon-rich part of the
[Si/Fe]-[Fe/H] locus.  Similarly, the lack of planet hosts in the
nickel-poor parts of the [Ni/Fe]-[Fe/H] locus suggests that planet hosts
might be nickel-enhanced as well.  Our hypothesis, therefore, is that
stars with planets have higher silicon-to-iron and nickel-to-iron ratios
than are typical of the field star population.  We also investigate a
possible relationship between planet detection and [Ti/Fe], even though
no such trend is obvious in Figure \ref{siscatter}.  Our aim is to
assess whether the visual evidence of Figure \ref{siscatter} corresponds
to statistically significant silicon and nickel enhancement among planet
hosts, and whether titanium abundance follows a similar trend, despite
the lack of an immediate visual relationship.

To test our hypothesis, we would like to measure the likelihood that
planet-host stars follow the same underlying [X/Fe] distribution as the
field-star population, where X = Si, Ni and Ti.  However, for any X,
[X/Fe] varies with [Fe/H], reflecting the history of Galactic chemical
evolution (Timmes, Woosley \& Weaver 1995).  Merely due to the
planet-metallicity correlation, planet hosts will assuredly follow a
different silicon or nickel abundance distribution than the field-star
population, as reflected in Figure 1 of \cite{bodaghee03}.  If element X
is intrinsically important to the process of planet formation, either as
a building-block of giant planets, or as a tracer of some other
fundamental process, the stars with the highest value of [X/Fe] {\it at
a given iron abundance} will be the most likely to harbor planets.  We
will use a Kolmogorov-Smirnov (K-S) test (see, e.g. Press et al. 1992)
to estimate the probability $k$\ that a synthetic set of stars with
planets, $B$, follows the same [Si/Fe], [Ni/Fe] and [Ti/Fe]
distributions as a control sample, $C$, of field stars, {\it if both
samples are drawn from the same iron abundance distribution.}  By
matching the [Fe/H] distributions of $B$ and $C$, we can control for the
effects of Galactic chemical evolution, ensuring that we investigate
only abundance patterns that correlate directly with the presence or
absence of giant planets.

At this point, we make no assumptions about the physical processes by
which planet formation might be stimulated by silicon or nickel
enrichment of the host star, nor the distributions of the variables that
control these processes (see \S \ref{model} for a simulation
of the relationship between [Si/Fe] and planet detection).  We therefore
choose a bootstrap Monte Carlo method, drawing from the SPOCS data with
replacement to find $B$ and $C$.  We perform 100,000 bootstrap
experiments to simulate the distribution $D_{\rm X}$ of
Kolmogorov-Smirnov probabilities that would be expected from bootstrap
realizations of a true probability ${\mathcal P}_{\rm X}$, that the
Galaxy's planet hosts and field stars follow the same [X/Fe]
distribution.  The bootstrap method will only tell us the form and
spread of $D_{\rm X}$---it cannot determine the median around which
$D_{\rm X}$\ would be distributed in the limit $k \rightarrow {\mathcal
P}_{\rm X}$.  If the abundance of X is correlated with probability of
planet detection, the probability $k$\ returned by the K-S test should
be low---but how low does it have to be to indicate a statistically
significant effect?  To answer this question, we need to test what
$D_{\rm X}$\ would look like if $k$\ were always calculated using two
samples identically distributed in [X/Fe].


Since we must control for the effects of Galactic chemical evolution, we
always calculate realizations of $k$\ using samples of planet hosts and
field stars selected from the same [Fe/H] distribution.  We can find the
K-S probability $q$ that planet hosts of set $B$ and field stars of set
$C$ follow the same distribution in [Fe/H]: By construction, the
underlying distribution ${\mathcal Q}$ of $q$ is unity.  In practice,
$q$ can never reach unity because of the impossibility of selecting
different two sets of stars with [Fe/H] distributions that exactly match
from a parent set of finite size.  $q > 0.5$ is consistent with the
hypothesis that $B$ and $C$ truly have matching [Fe/H] distributions.
The simulated distribution of $q$, $D_{\rm Fe}$, can be used as a
benchmark to assess $D_{\rm X}$.  We estimate ${\mathcal P}_{\rm X}$,
the true probability that planet hosts in the SPOCS data follow the same
[X/Fe] distribution as other field stars of the same metallicity, by
the following formula:
\begin{equation}
{\mathcal P}_{\rm S} = {\int_0^1 D_{\rm X} \,  D_{\rm Fe} \,
p \, dp \over \int_0^1 D_{\rm Fe}^2 \, p \, dp},
\label{probability}
\end{equation}
where $p$ denotes probability.  See Appendix \ref{appendix1}
for a detailed description of the statistical methods used to find
$D_{\rm X}$ and $D_{\rm Fe}$.

\begin{figure}
\plotone{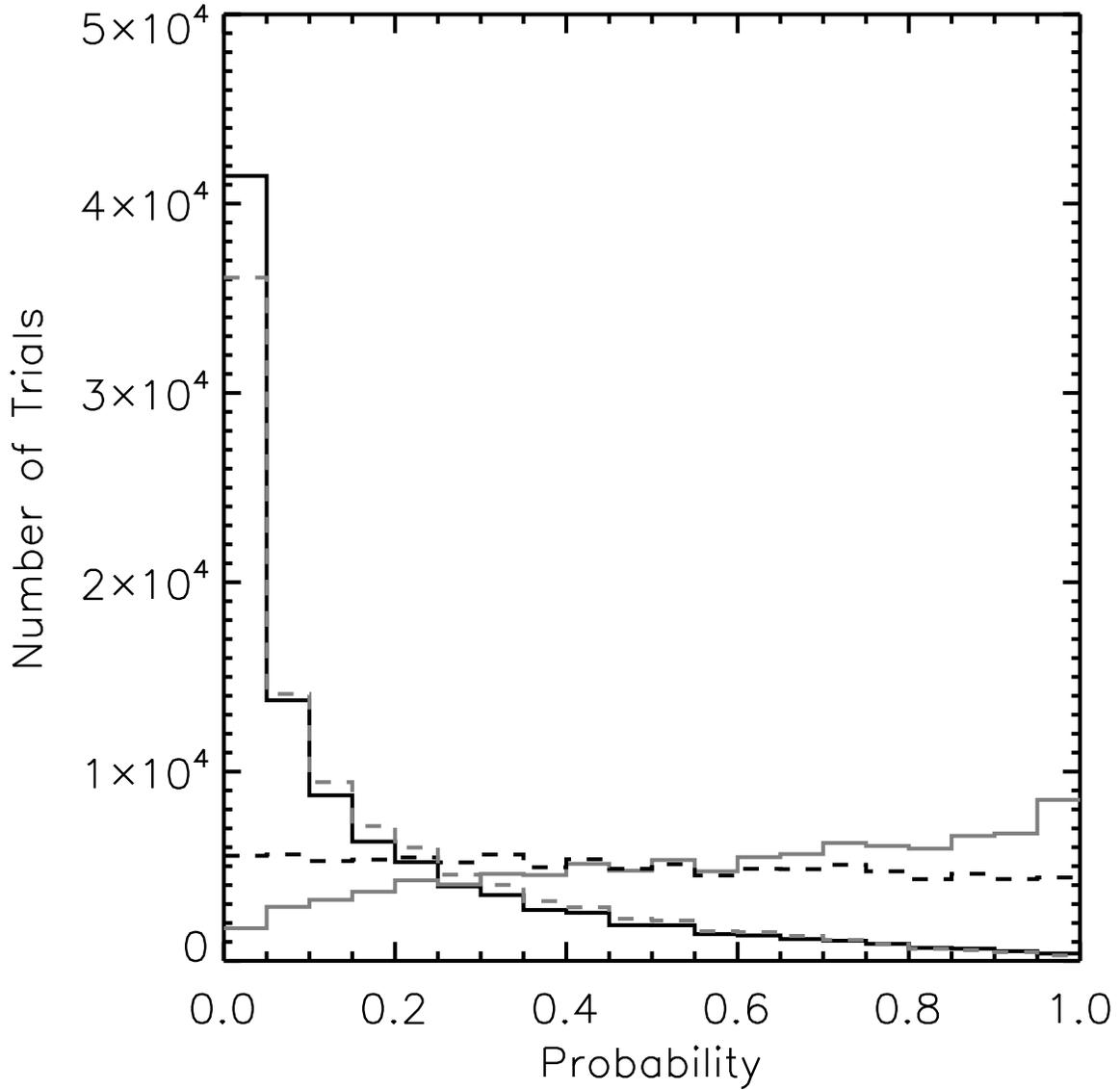}
\caption{Histogram of probabilities, returned by the K-S test, that the
planet hosts are from the same abundance distribution as the control set
of field stars.  The test results for [Si/Fe] are plotted in solid
black, [Ti/Fe] in dash-dot black, [Ni/Fe] in dash-dot gray, and [Fe/H]
in solid gray.}
\label{ksprob}
\end{figure}

Figure \ref{ksprob} shows the result of the Monte Carlo simulations.
For silicon, we find a probability ${\mathcal P}_{\rm Si}$ that planet
hosts and field stars are drawn from the same [Si/Fe] distribution of
0.23.  In addition, $D_{\rm Si}$ clearly has a different form than
$D_{\rm Fe}$.  Forcing the control set to follow the planet hosts'
metallicity distribution succeeds in almost all simulations: The median
of $D_{\rm Fe}$\ is 0.61 and $D_{\rm Fe}$ falls as $q \rightarrow 0$.
However, $D_{\rm Si}$\ has median 0.079, and declines rapidly as $p
\rightarrow 1$.  The results shown in Figure \ref{ksprob} indicate that
even when the effects of Galactic chemical evolution are taken into
account, stars with planets do not have the same silicon abundance
distribution as the general field star population.  There is, therefore,
evidence that, at a given iron abundance, a star's probability of
harboring a detectable planet depends on its silicon abundance.

For nickel, we find ${\mathcal P}_{\rm Ni} = 0.25$, and $D_{\rm Ni}$,
like $D_{\rm Si}$, is skewed toward low probabilities that the planet
hosts and field stars have matching [Ni/Fe] distributions.  There is
then evidence that, at a given [Fe/H], a star's probability of harboring
a detectable planet depends on its nickel abundance.  For titanium, we
find ${\mathcal P}_{\rm Ti} = 0.78$, and $D_{\rm Ti}$ appears to have a
similar form to $D_{\rm Fe}$.  There is, therefore, no statistical
evidence that titanium abundance is correlated with planet detection
probability, other than as an overall metallicity indicator.


We still have not given direct evidence for our original hypothesis,
that planets are more likely to be detected around silicon-rich and
nickel-rich, than silicon-poor and nickel-poor stars---we have merely
shown that the silicon abundances of planet hosts are systematically
{\it different} from those of the field-star population.  In Figure
\ref{siscatter}, it looks like stars with planets might be more likely
to be silicon- and nickel-enhanced.  This impression is confirmed by
Figure \ref{sifehists}, where we have binned the SPOCS data by iron
metallicity and plotted the percent of stars with planets as a function
of [Si/Fe] and [Ni/Fe].  Two things are apparent in Figure
\ref{sifehists}: (1) the stars with planets are to be found at the top
of the range of silicon or nickel abundances exhibited by the stars in
each bin; and (2) the general trend is that the fraction of stars with
planets rises toward higher silicon or nickel abundance.  The exception
to (2) is the bin $0.35 < {\rm [Fe/H]} < 0.45$\ dex---there are only 33
stars in this bin (as opposed to more than 100 in each of the other
bins), so it may suffer from small-sample statistics.

We note that accurate calculation of ${\mathcal P}_{\rm Ni}$ and
${\mathcal P}_{\rm Si}$ depends on the stars in SPOCS having been
selected for the Keck planet-search program without any reference to
silicon or nickel content.  One of program's aims was to obtain
high-resolution spectra of as many planet hosts as possible, so the host
stars of planets discovered by other groups were added to the survey.
This definitely introduces a bias in the planet hosts' [Fe/H]
distribution, which we have corrected by comparing the planet hosts with
control samples that have the same [Fe/H] distribution.  If, as our
analysis suggests, planet hosts are more likely to be silicon- or
nickel-enhanced than field stars of the same metallicity, adding planet
hosts discovered by other groups to the SPOCS survey would amplify any
existing correlation between presence of planets and [Si/Fe] or [Ni/Fe].

Recognizing that observing planet hosts discovered by other groups would
bias the [Fe/H] distribution of SPOCS, \cite{FV05} identified a set of
stars with uniform planet detectability, requiring that all stars in
this set be independently selected targets for the Keck, Lick and AAT
planet searches, and have at least 10 observations spanning four years
with 30 m s$^{-1}$ or better radial-velocity precision.  This uniform
set, which contains 850 stars and 47 planet hosts, was used to study the
planet-metallicity correlation.  To confirm that the suggestion of
silicon and nickel enhancement among planet hosts is not the result of
SPOCS selection biases, we perform our statistical analysis of the
[Si/Fe], [Ni/Fe] and [Ti/Fe] distributions of planet hosts on the
uniform set, using 10,000 bootstrap Monte Carlo simulations.  We find
${\mathcal P}_{\rm Si} = 0.20$, ${\mathcal P}_{\rm Ni} = 0.25$, and
${\mathcal P}_{\rm Ti} = 0.94$.  The value of ${\mathcal P}_{\rm Si}$ is
slightly lower using the uniform data set than the entire SPOCS
compilation, and the value of ${\mathcal P}_{\rm Ni}$ is identical.
This shows that our finding that planet hosts have enhanced silicon
content was not the result of a selection effect.  Here, the [Ti/Fe]
distribution of the planet hosts is found to match that of the control
set even more closely than when the experiment was performed on the
entire SPOCS dataset, which underscores our finding that, for a given
value of [Fe/H], titanium abundance is not correlated with the
probability of finding a planet.



\section{Empirical Model of Planet-[Si/Fe] Correlation}
\label{expsim}

\begin{figure}
\plottwo{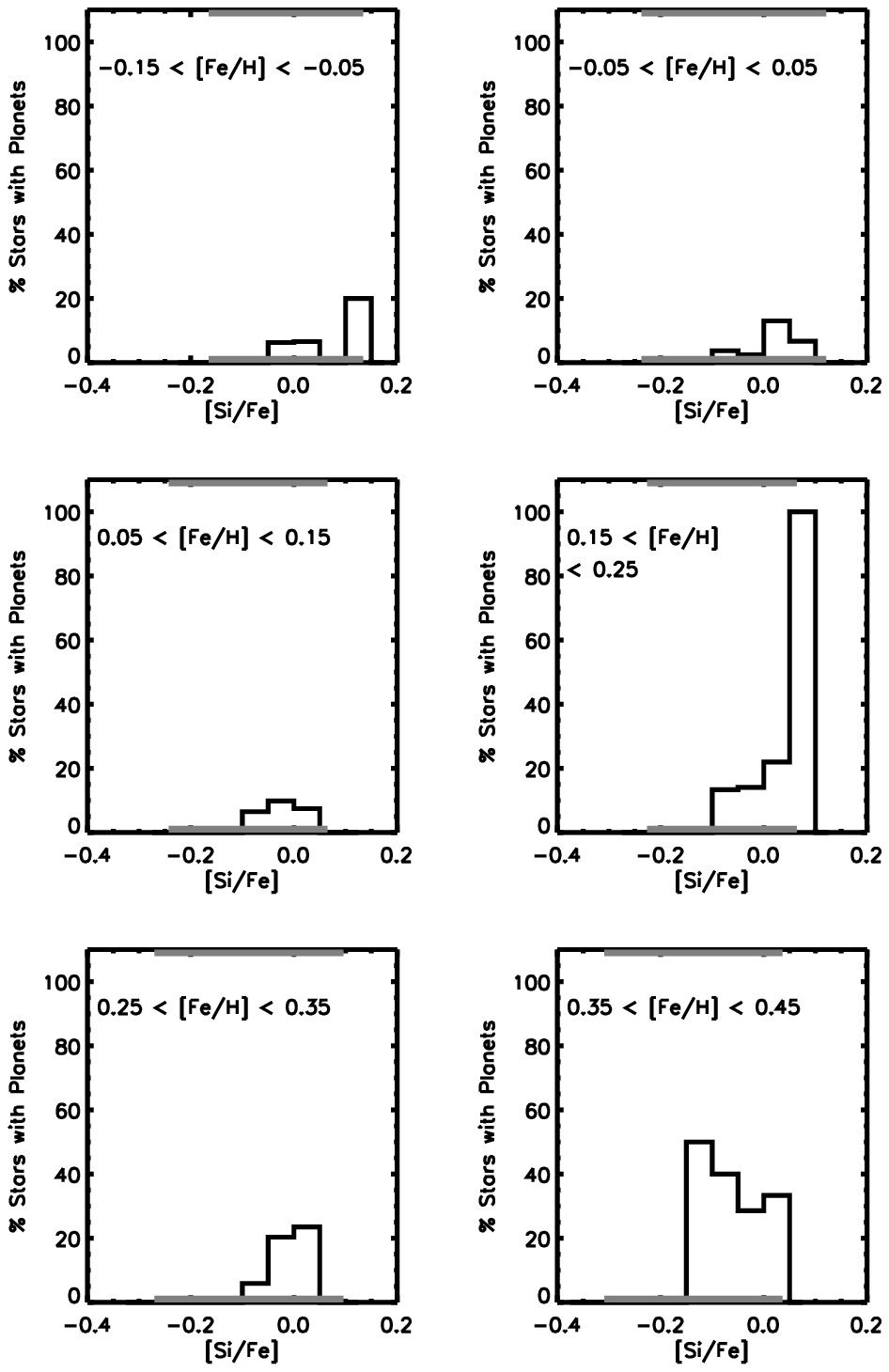}{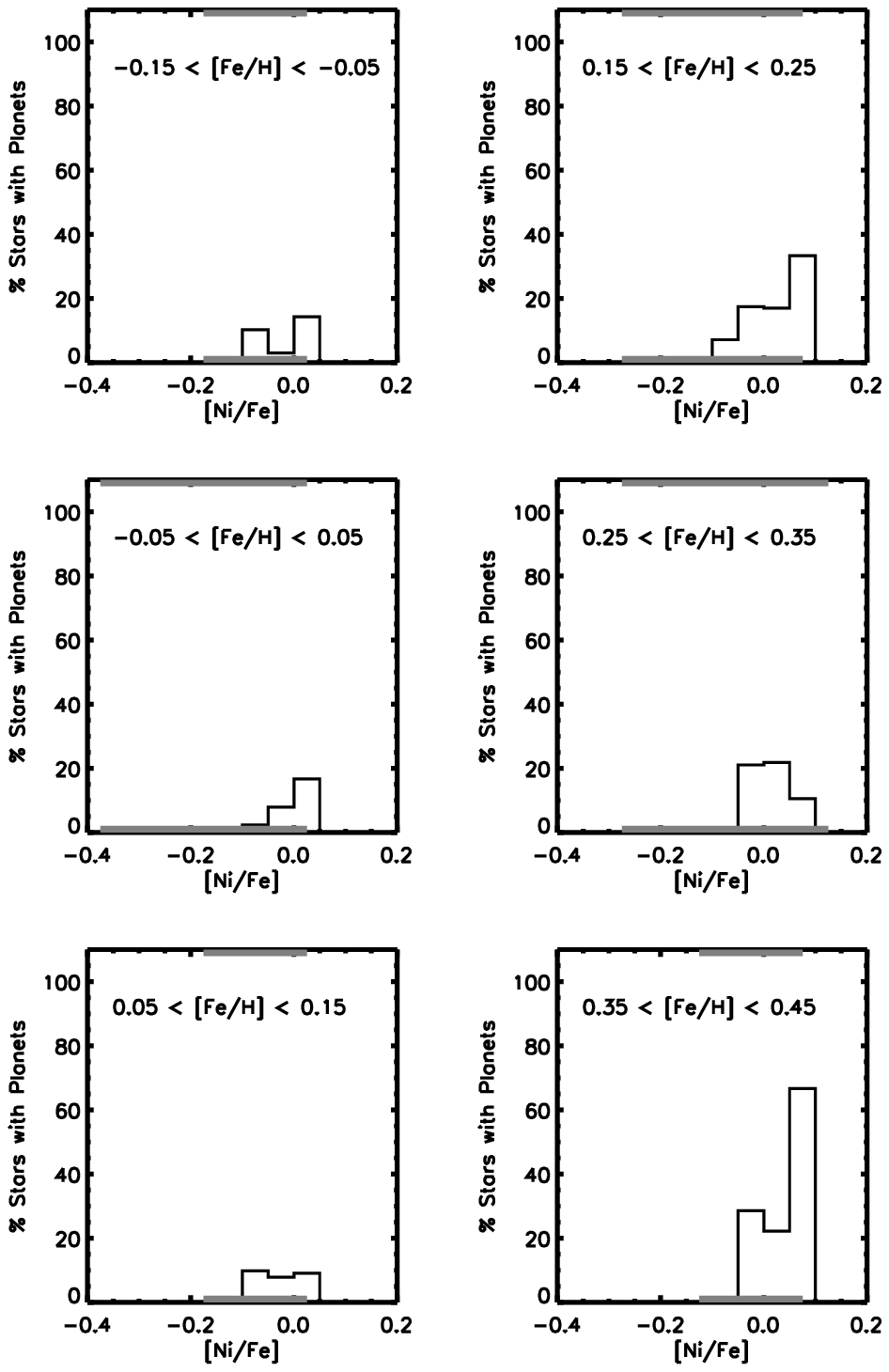}
\caption{Percent of stars with planets as a function of [Si/Fe] (left)
and [Ni/Fe] (right) for different bins in [Fe/H].  The gray bars on the
bottom and top of the plot box show the range of values of [X/Fe]
present in SPOCS in each individual [Fe/H] bin.  Stars with planets
always come from the top of the field-star [Si/Fe] and [Ni/Fe] ranges.}
\label{sifehists}
\end{figure}

Having determined that planet hosts show evidence of silicon and nickel
enhancement, we would now like to see if it is possible to
empirically quantify the relationship between these abundances and
probability of planet detection.  Examination of Figure \ref{sifehists}
reveals that once the planet hosts of SPOCS are binned according to iron
abundance, small-sample statistics prevent us from robustly determining
a planet-silicon or planet-nickel correlation.  Any number of different
functions could be used to fit the histograms in Figure \ref{sifehists},
and the best fit would be noticeably different in each iron-abundance
bin.  While it is certainly possible, even likely, that the relationship
between probability of planet detection and silicon/nickel abundance
changes with overall metallicity, it is not within our power to explore
this empirically until more planet hosts have been discovered and
surveyed.

We will use silicon as the test element to see whether or not the we can
empirically determine the relationship between probability of planet
detection and [X/Fe] in the SPOCS data.
We choose to test the simplest hypothesis: at a given
metallicity, the probability of planet detection follows a power law in
silicon abundance.  We seek a model similar to the planet-metallicity
correlation presented by \cite{FV05}, which uses the iron abundance:
\begin{equation}
P({\rm planet}) = 0.03 \times \left [ { N_{\rm Fe} / N_{\rm H} \over
\left ( N_{\rm Fe} / N_{\rm H} \right )_{\odot} } \right ]^2  = 0.03
\times 10^{2.0{\rm [Fe/H]}} .
\label{pmceq}
\end{equation}
We therefore assume the following form for the planet-silicon
correlation:
\begin{equation}
\left [ P({\rm planet}) \right ]_{\rm [Fe/H]} \propto
10^{b \, {\rm [Si/Fe]}} ,
\label{psieq}
\end{equation}
where the notation $[Y]_{\rm [Fe/H]}$ denotes the quantity $Y$ measured
at constant [Fe/H].  The proportionality constant in equation
\ref{psieq} will be allowed to change with [Fe/H], but the exponent $b$
will not.  We want to estimate the value of $b$ implied by the SPOCS
data set.

To do this, we again draw bootstrap realizations of the SPOCS data set,
but we do not identify the actual planet hosts in each realization.
Instead, we assign planets based on a two-dimensional probability
distribution:
\begin{equation}
P({\rm planet}) = {\mathcal F}({\rm [Fe/H],[Si/Fe]}) .
\label{fform}
\end{equation}
${\mathcal F}$ is a two-dimensional histogram constructed such that each
cut at constant [Fe/H] obeys equation \ref{psieq}.  The proportionality
constant for each [Fe/H] cut is set so that the planet-metallicity
correlation remains intact:
\begin{equation}
\sum_{\rm [Si/Fe]} {\mathcal F} = 0.03
\times 10^{2.0 {\rm [Fe/H]}} .
\label{norm}
\end{equation}
We assume that each realization of the SPOCS data samples the entire
range of [Si/Fe] values present in metal-rich stars, so equation
\ref{norm} performs a discrete normalization of ${\mathcal F}$, such
that ${\mathcal F}$ is zero in regions of the [Si/Fe]-[Fe/H] plane that
contain no SPOCS stars.  Using ${\mathcal F}$, we assign planets to the
stars in each bootstrap realization of SPOCS, to simulate the
relationship between planet detection and [Si/Fe] that would be present
in SPOCS, if $b$ were the proper exponent for equation \ref{psieq}.


As in \S \ref{observed}, we create a control set with the same [Fe/H]
distribution as the set of synthetic planet hosts, and use the K-S test
to check for differences between the two sets' [Si/Fe] distributions.
We perform 10, 10,000-simulation Monte-Carlo experiments, each with a
different exponent $b$ in equation \ref{psieq} determining the form of
${\mathcal F}$.  For each $b$, the Monte-Carlo simulations will tell us
the distribution of probability that the set of synthetic planet hosts
has the same silicon-abundance pattern as field stars with the same
metallicity.  We will call this distribution $H$.  By comparing $H$ with
the previously determined distribution $D_{\rm Si}$, we can find the
exponent $b$ that best models the planet-silicon correlation, as present
in the SPOCS data.  See Appendix \ref{appendix2} for a detailed
description of the statistical methods used in these simulations.

\begin{figure}
\epsscale{1.0}
\plotone{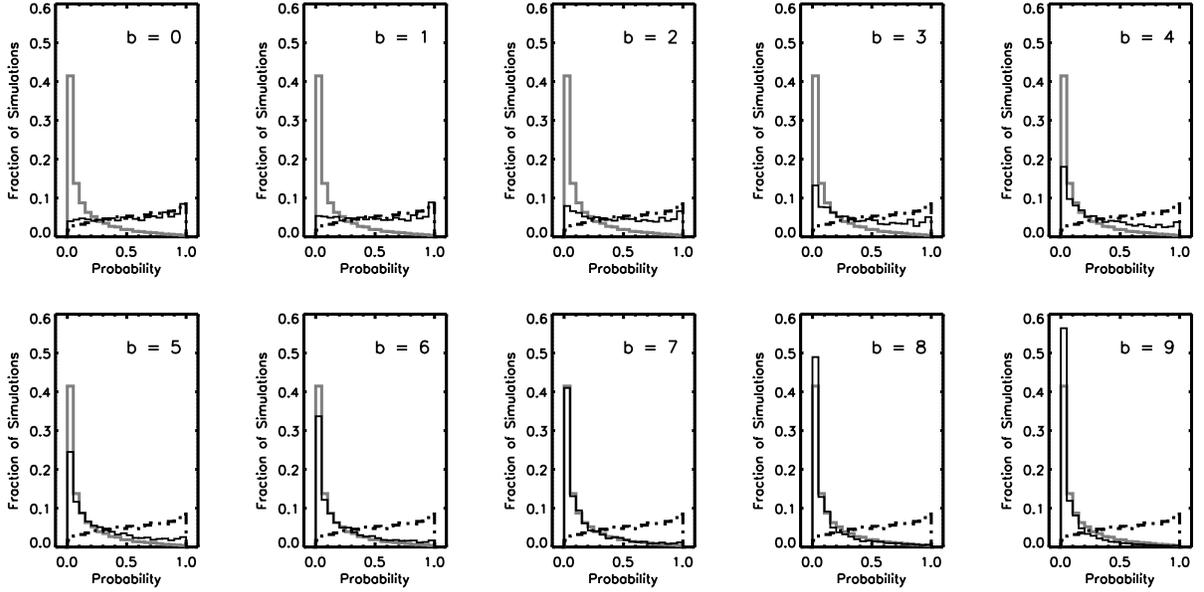}
\caption{Results of simulations to determine power-law exponent
governing the planet-silicon relation.  In solid black are histograms of
probability that planet hosts are drawn from the same silicon-abundance
distribution as the field stars of the same iron abundances, if planet
frequency follows a power law with exponent $b$.  In each panel, the
distribution $D_{\rm Si}$, the probability that the actual planet hosts
in SPOCS have the same silicon-abundance pattern as the field-star
population, is shown in solid gray.  For reference, $D_{\rm Fe}$, the
histogram of probabilities that sets of stars selected by iron abundance
do, in fact, have matching [Fe/H] distributions, is plotted in
dash-dotted black.}
\label{slopesim}
\end{figure}

Figure \ref{slopesim} shows the results of this experiment.  Each panel
shows $H$ for a particular power-law exponent $b$ that governs the
planet-silicon correlation.  Also plotted for comparison are $D_{\rm
Si}$ and $D_{\rm Fe}$, as shown in Figure \ref{ksprob}.  For $b = 0$,
the null hypothesis in which likelihood of planet detection does not
depend on silicon abundance except as it traces iron abundance (or
overall metallicity), $H$ closely follows the form of $D_{\rm Fe}$,
indicating that, as expected, the synthetic planet hosts and control
stars with matching iron abundances follow the same [Si/Fe]
distribution.  As $b$ increases, $H$ gets more and more skewed toward
low probabilities.  The best match between $H$ and $D_{\rm Si}$,
calculated by finding the value of $b$ that corresponds to the maximum
overlapping area under $H$ and $D_{\rm Si}$, is obtained with $b = 7$,
so according to this analysis, the planet-silicon correlation has the
form
\begin{equation}
\left [ P({\rm planet}) \right ]_{\rm [Fe/H]} \propto 10^{7\, {\rm
[Si/Fe]}} .
\label{psicorr}
\end{equation}
Since [Si/Fe] = [Si/H] - [Fe/H], equation \ref{psicorr} can be
rewritten as
\begin{equation}
\left [ P({\rm planet}) \right ]_{\rm [Fe/H]} \propto 10^{7 \,
{\rm [Si/H]}} .
\label{psihcorr}
\end{equation}

Such a strong correlation between silicon abundance and planet detection
suggests an absolutely remarkable physical importance of silicon in the
planet-formation process.  It is unlikely, however, that silicon really
is such a strong accelerant of planet formation, as a power-law exponent
of seven indicates.  Indeed, in Figure \ref{slopesim}, varying $b$
between 4 and 10 makes little difference in the form of $H$, which we
take as evidence that either (1) a power law is not the correct form for
the planet-silicon correlation, or (2) the SPOCS data do not robustly
determine the relationships between probability of planet detection and
[X/Fe].  No doubt such a strong difference in $b$ would be obvious if
stars at a given [Fe/H] could span a wide range in [Si/Fe], but this is
not the case, as we attach planets to observed values of [Si/Fe].   The
sharper the planet-silicon correlation, the more the planet hosts merely
get pushed to the top of the [Si/Fe] vs.  [Fe/H] locus of Figure
\ref{siscatter}, until there are no higher values of [Si/Fe] available.

Rather than a power-law planet-silicon correlation with a large
exponent, we suspect a threshold phenomenon: All circumstellar disks
that can form Jovian planets do so, and the division between disks that
can form planets and those that cannot is a sharp cutoff, a step
function in one parameter only.  That one parameter may be silicon
abundance itself, or it may be another physical quantity of which
silicon is an indicator.  If there were cutoff in silicon abundance,
above which planets would definitely form and below which they could
not, Figure \ref{sifehists} would show a step function in every panel.
Instead, the step function is smeared out, so planets appear merely more
frequently, rather than exclusively, around metal-rich and/or
silicon-rich stars.  Planet hosts and stars without planets occupy
overlapping regions of the [Fe/H]-[Si/Fe] plane, suggesting that stars
with high metallicity and high silicon abundance are more likely, but
not guaranteed, to have protostellar disks that meet the criterion for
planet formation.  The exact value of the minimum silicon abundance
required for planet formation must depend on the characteristics of the
individual star-disk system.  In the next section, we describe Monte
Carlo simulations that predict frequency of planet formation based on
iron abundance, silicon abundance, disk mass and radii, and disk
lifetime.  We will use a different method than the power-law fit
presented in this section to compare our simulation results with the
observed planet-silicon correlation.

\section{Probability of Planet Detection: Monte Carlo Simulations}
\label{model}

In the core accretion model of planet formation, the critical quantity
that controls whether gas giant planet formation can proceed is the
solid surface density in the protostellar disk.  A Jupiter-type planet
will form if the disk has a sufficient concentration of solids just
beyond the ice line for a protoplanetary core to reach runaway gas
accretion before the disk dissipates (Pollack et al. 1996).  If the disk
solid surface density in the protoplanet's feeding zone is too low for
runaway gas accretion to begin, a Neptune-mass planet will form.
According to the calculations by Hubickyj, Bodenheimer \& Lissauer (in
press), a Jupiter-mass planet can form in 2.3 Myr in a disk with solid
surface density $\sigma_{\rm solid} = 10 \; {\rm g \, cm^{-2}}$ at 5 AU,
whereas decreasing the surface density to $\sigma_{\rm solid} = 6 \;
{\rm g \, cm^{-2}}$ increases the timescale for Jupiter formation to
13.3 Myr.  Thus, a $40\%$ decrease in solid surface density just beyond
the ice line leads to a nearly $500\%$ increase in Jupiter's formation
time, indicating a sharp threshold in the solid surface density required
for giant planet formation in the Solar nebula.  A disk enriched in
either silicon or nickel would be more likely than its unenriched
counterpart to be over the solid surface density threshold for planet
formation.

The planet-metallicity correlation, modeled using the SPOCS data by
\cite{FV05}, is a natural consequence of the core accretion theory of
planet formation, since all solid species are metals.  It has not yet
been established whether [Fe/H] is a good predictor of planet
detectability merely because it traces the total metal content of the
star (i.e., all metals are equally useful at forming planets), or
because iron is a particularly important core-forming material.  In the
simplest interpretation of core accretion theory, the importance of each
element to planet formation is in direct proportion to its abundance.
The same multiplicative factor of enhancement over solar abundance adds
the most solid mass when the enhanced element accounts for a large mass
fraction of protoplanetary material.  Correspondingly, depleting an
otherwise metal-rich disk of a naturally abundant solid, such as oxygen,
would seriously damage its chances for giant planet formation.

To illustrate the effect of different element enhancements on solid
surface density, consider a disk with solar abundances of every element
except iron.  The disk has iron abundance double that of the sun
([Fe/H] = 0.3), and its solid surface density is $12\%$ higher than a
disk with the same mass and inner and outer radii, but with Solar iron
abundance (we use the solar abundances reported by Anders \& Grevesse
[1989], normalized to $\log N_{\rm H} = 12.00$).  A disk with solar
composition except for a factor of 2 enhancement of silicon, such that
$\log N_{\rm Si} = 7.55 + 0.3$, has solid surface density $6.7\%$ higher
than a disk with the same mass and radii, with solar composition.
Doubling the abundance of nickel, $\log N_{\rm Ni} = 6.25 + 0.3$, gives
only a $0.7\%$ increase in $\sigma_{\rm solid}$.  If the oxygen
abundance of a disk with solar composition were doubled, keeping the
same mass and radii, the solid surface density at the ice line would
increase by $56\%$.  Assuming carbon is present mainly in gaseous CO and
CH$_4$ (Lewis \& Prinn 1980), oxygen, iron and silicon are the elements
that contribute the most solid mass to planet formation.  The
contribution of nickel, due to its low abundance, is much less
significant.  Galileo observations of Callisto and Ganymede, which
formed outside the snow line of the proto-Jovian nebula, suggest that
both are composed of $\sim 50 \%$ ice by mass (Sohl et al.  2002).
Jupiter probably could not have formed in an oxygen-poor, yet iron-rich
circumstellar disk.


We propose that there are two reasons why silicon abundance is
correlated with planet detection: (1) silica and silicates form a
significant portion of the grains that seed the planet formation
process, silicon being the third most abundant solid element by mass at
Jupiter's position in the Solar nebula, and (2) most importantly,
because silicon and oxygen are both $\alpha$-elements, the Si/O
abundance ratio is roughly constant among metal-rich stars.  Therefore,
[Si/Fe] is a tracer of [O/Fe], and silicon enrichment implies a star is
oxygen-rich as well.

The link between silicon and oxygen abundance, and the preponderance of
oxygen in protoplanetary material, prompted us to perform Monte Carlo
simulations of the observed planet-silicon correlation.  Our simulation
combines observationally determined properties of protostellar disks,
with new simulations, using the core accretion model, of
planet-formation timescale as a function of solid surface density.  We
generate synthetic star-disk systems described by the independent random
variables disk mass $M_d$, outer radius $r_{out}$, disk lifetime $T$,
[Fe/H] and [Si/Fe].  We do not simulate the effect of nickel enhancement
on likelihood of planet formation.  The solid surface density at
Jupiter's location in each disk is calculated from the disk mass and
chemical composition.  If the solid surface density is high enough for a
protoplanetary core to reach runaway gas accretion within the disk
lifetime, the star becomes a planet host.  We generate 100,000 star-disk
systems, which we use to quantify the relationship between probability
of planet detection and silicon abundance.  Our results should reproduce
both the planet-metallicity correlation and the trend reported in this
paper, of planets being found preferentially around silicon-rich stars.

In \S \ref{ssd}, we give the results of new numerical simulations that
model the relationship between disk solid surface density and timescale
for planet formation.  In \S \ref{distribs}, we describe the random
variables that specify star-disk systems in our Monte Carlo simulations.
Finally, in \S \ref{modelresults}, we discuss our simulations'
predictions about $\alpha$-enrichment in planet hosts, and compare these
predictions with the trends present in the SPOCS data.

\subsection{Solid Surface Density and Timescale for Planet Formation}
\label{ssd}

We use the theoretical model of planet formation described by Laughlin,
Bodenheimer \& Adams (2004, hereafter LBA) to calculate planet formation
timescale as a function of solid surface density.  Initially, a
protoplanetary core of mass $M_{\earth}$ is embedded at 5.2 AU in a disk
of age $10^5$ years, surrounding a T-Tauri star of mass $1 M_{\sun}$.
This is a good match for the SPOCS data, which, consisting of FGK stars,
have a median stellar mass of $1.14 M_{\sun}$.  The disk is flat and
passive, and isothermal in the vertical direction.  The stellar
effective temperature $T_*(t,M_*)$ and luminosity $L_*(t,M_*)$ are
adopted from published pre--main-sequence stellar evolution tracks
(D'Antona \& Mazzitelli 1994).

The contraction and buildup of protoplanetary cores and their gaseous
envelopes embedded in our model evolving disk are computed with a
Henyey-type code (Henyey et al. 1964).  Following the argument of
\cite{podolak03} that grain settling in the protoplanetary envelope
would reduce envelope opacity where grains exits, we adopt grain
opacities of $\sim 2\%$ of the interstellar values used in
\cite{pollack96}.  We use a core accretion rate of the form $dM_{\rm
core}/dt = C_1 \pi \sigma_{\rm solid} R_c R_h \Omega$ (Papaloizou \&
Terquem 1999), where $\sigma_{\rm solid}$ is the surface density of
solid material in the disk, $\Omega$ is the orbital frequency at 5.2 AU,
$R_c$ is the effective capture radius of the protoplanet for solid
particles, $R_h = a[M_{\rm planet} / (3 M_*)]^{1/3}$ is the tidal radius
of the protoplanet (where $a$ is the semimajor axis of the protoplanet's
orbit), and $C_1$ is a constant near unity.  The outer boundary
conditions for the protoplanet include the decrease with time in the
background nebular density and temperature.

One parameter in the model is the gas/solid ratio just beyond the ice
line in the Solar nebula, at the onset of protoplanetary core formation.
To find the gas/solid ratio, we assume the Solar elemental abundances of
\cite{standardsolar} and follow the chemical model of Hersant, Gautier
\& Lunine (2004, hereafter HGL).  As in \cite{lp80}, we assume all
carbon in the pre-solar nebula was in CO and CH$_4$, all oxygen was in
CO and H$_2$O, and all nitrogen in N$_2$ and NH$_3$.  If the CO/CH$_4$
ratio in the pre-solar cloud was preserved in the Solar nebula, CO/CH$_4
= 10$.  HGL present stability curves for the CHON ices, which show that
CO at 5 AU does not freeze until $\sim 3$ Myr after disk formation, nor
become trapped in clathrate hydrates until 1.6 Myr after disk formation,
by which the protoplanet has already started to build up a gaseous
envelope (LBA).  CH$_4$ freezes at 2.8 Myr and clathrates at 1.3 Myr, by
which time $\sim 1/2$ of the core mass has already accumulated.  The
only volatile besides H$_2$O that could possibly freeze or clathrate
while the solid core is truly embryonic is NH$_3$ at 0.9 Myr.  However,
at N$_2$/NH$_3$ = 10 (Lewis \& Prinn 1980, Irvine \& Knacke 1989),
NH$_3$ comprises such a minor proportion of the CNO-bearing molecules
that it cannot significantly speed up the formation of protoplanetary
cores.

We therefore assume that all CNO species except water are in the gas
phase.  Although carbon can form refractory organic compounds, these
should be present in small amounts compared to CO (see, however, Lodders
2004).  Considering elements with abundances $10^6$ and above (relative
to $N_H = 10^{12}$), we assume solid Na, Mg, Al, Si, Ca, Fe and Ni.  We
also assume solid sulfur, in the form of FeS and H$_2$S.  According to
\cite{pasek05}, the condensation front of troilite (FeS) is between 1
and 2 AU from the Sun at the start of our simulations, $10^5$ years
after Solar-nebula formation.  H$_2$S is half frozen at the snow line,
and continues to solidify as the disk evolves.  Two hydrogen atoms for
every oxygen bound in H$_2$O are frozen as well, and He, C, N, Ne and Ar
are entirely in the gas phase.  At Solar abundances and with CO/CH$_4$ =
10, 62\% of the oxygen atoms are frozen in H$_2$O, and the rest are in
CO gas.  This gives an initial gas/solid ratio just beyond the ice line
in the Solar nebula of $G/S = 100$, adjusted from $G/S = 70$ in
LBA.  The model accounts for the decrease in $G/S$ with time as gas
evaporates from the disk using the same functional form as LBA.

\begin{figure}
\plotone{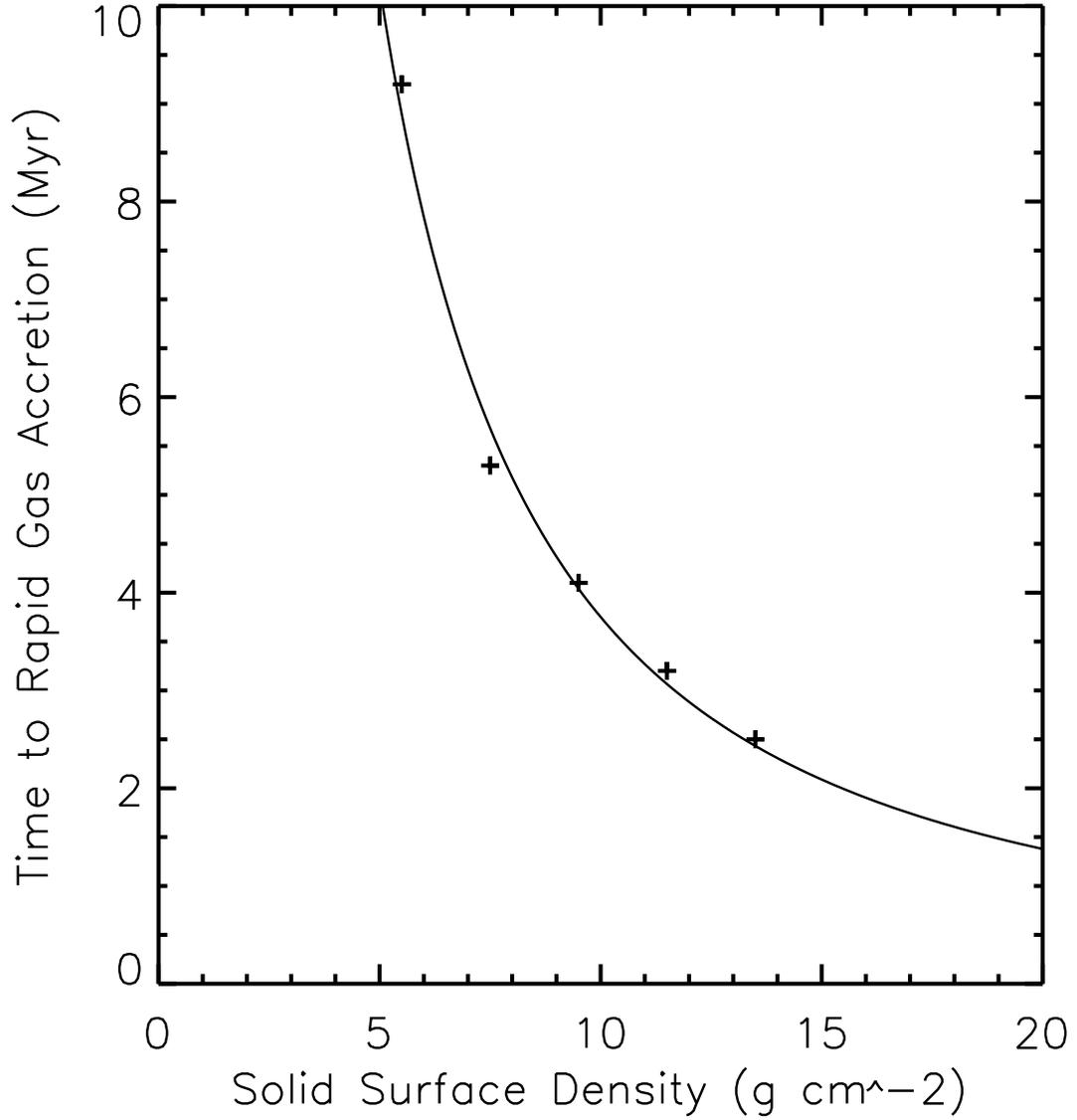}
\caption{Time till gas accretion of a Jovian protoplanet, as a function
of solid surface density, at 5.2 AU from the young Sun.  A power law,
$t_{\rm rga}(\sigma)$, is plotted over points showing the
results of five simulations.}
\label{formtime}
\end{figure}

We calculated the time to runaway gas accretion for different initial
values of solid surface density.  At $\sigma_{\rm solid} = 5.5, 7.5,
9.5, 11.5$ and 13.5 g cm$^{-2}$, the times to accelerating accretion are
9.2, 5.3, 4.1, 3.2 and 2.5 Myr, respectively.  To describe planet
formation time as a function of surface density, we fit an power law
to the simulation results:
\begin{equation}
t_{\rm rga} = \left ( {\sigma_{\rm solid} \over 25.0 {\rm g \: cm}^{-2}}
\right )^ {-1.44}
\label{formtimeeq}
\end{equation}
where $t_{\rm rga}$ is the time to rapid gas accretion in Myr.  The
simulation results and the fit in equation \ref{formtimeeq} are plotted
in Figure \ref{formtime}.  For the synthetic star-disk systems in our
Monte Carlo simulation, if $t_{\rm rga}$ is less than the disk lifetime,
a giant planet forms.


\subsection{Disk Properties}
\label{distribs}

Since surface density and disk lifetime together determine whether
Jovian planets can form, we seek to describe these in terms of
independent, random variables that can be the basis of Monte Carlo
simulations.  If one assumes (1) that disk lifetime does not depend on
disk mass, both of these can be chosen as random variables in the
simulations.  We have no observational information about what the
surface density 5 AU from protostars might be, so we make the further
assumption (2) that, although the outer disk radius may vary, the disks
always have the same inner radius and surface density distribution
(gas+solid) is always characterized by the same power law in radius.  If
we assume (3) abundances in stars are the same as in their circumstellar
disks, and (4) that the [X/Fe] ratios of all $\alpha$-elements are
roughly the same in all metal-rich stars, such that [Si/Fe] =
[$\alpha$/Fe], we can specify a star-disk system with two more random
variables that have observational constraints: [Fe/H] and [Si/Fe].

Observations of IC 348 by \cite{hll01a} indicate that disk lifetime in
the cluster may depend on the spectral type of the star: No
circumstellar disks were detected around OBAF T-Tauri stars, suggesting
that these disks have shorter lifetimes than the disks surrounding
late-type stars.  Statistically, this would lead to shorter lifetimes
for higher-mass disks, since disk mass is directly correlated with star
mass (Brice\~{n}o et al. 2001).  However, no difference in disk
frequency between G and K T-Tauri stars was detected by \cite{hll01a},
nor was the ratio of (accreting) CTTSs, to (presumed diskless) WTTSs in
Taurus-Auriga found to vary with age (Kenyon \& Hartmann 1995). We wish
to simulate the frequency of planet formation by late F, G and K dwarfs,
since more massive stars are excluded from radial-velocity planet
searches, including SPOCS.  Therefore, we conclude that, within the
domain of our simulation, assumption (1) is justified.


Assumption (3) is implicit because we are using the silicon abundance
observed by SPOCS in evolved, main-sequence stars to determine whether
conditions in a star's accretion disk were right for planet formation.
Our simulation assumes that disks do not fractionate in element abundances,
at least within 5.2 AU of their protostar.  We must also assume that the
silicon initially present in a star's photosphere does not significantly
settle inward during the star's main-sequence lifetime.  However, the
agreement between photospheric and meteoritic silicon abundance in the
Solar system (Anders \& Grevesse 1989) indicates that stellar silicon
abundance can be used to diagnose conditions in planet-forming disks.

Assumption (4) is the basis for the assertion that disk mass, [Si/Fe],
and [Fe/H] fully specify the amount of solid material available for
planet formation.  We assume that the $\alpha$-elements are present in
the same proportions in all metal-rich stars.  The most important
$\alpha$-element in our simulation is oxygen, since it is the most
abundant solid material in planet-forming disks.  \cite{abtrends} find
that the $\alpha$-elements Mg, Si, Ti and Ca in thin- and thick-disk
stars show similar behavior in the [X/Fe] vs. [Fe/H] plane, with [X/Fe]
flattening and staying relatively constant above solar metallicity.
However, they find that oxygen shows a decreasing trend at high
metallicity.  \cite{abtrends2} also find that [O/Fe] continues to
decline at ${\rm [Fe/H]} \geq 0.0$, which does not match the trend in
[Si/Fe] either in their own data or in the SPOCS data (see Figure
\ref{siscatter}).  However, the models of \cite{ofetheory} predict
[O/Fe] flattening at solar metallicity, which is supported by the
observations of \cite{abtrends3}.  It appears that the evolution of
[O/Fe] at high metallicity is not fully understood, but as long as we
may at least say that silicon traces oxygen and the other
$\alpha$-elements {\it better} than iron does, we have a basis for
assumption (4).

We argue, therefore, that a planet-forming star-disk system can be
described by five random variables: disk mass, outer radius, disk
lifetime, [Fe/H] and [Si/Fe].  \S \ref{mass}, \ref{lifetime} and
\ref{comp} describe the observationally motivated distributions from
which these random variables are chosen.

\subsubsection{Disk Mass, Radius and Surface Density Profile}
\label{mass}

To characterize the distribution of disk masses around pre-Solar stars,
we first assume that a large number of independent processes determine
the disk mass as the star evolves toward the T-Tauri phase.  For
example, the behavior of protostellar collimated outflows, which largely
control the envelope dissipation rate during the Class I phase, depends
on the strength of the magnetic field threading the disk (see, e.g., Shu
et al.  1995) and the accretion rate onto the star (Calvet 2003).  If
sufficiently many of the processes that control disk mass at the start
of planet formation are described by independent random variables, the
disk mass distribution naturally assumes a lognormal form.  This
characterization follows the same reasoning as \cite{clt}, who invoked
the central limit theorem to model the stellar IMF as a lognormal
distribution.

The fiducial point of the disk mass distribution is determined from the
observations of \cite{diskmass}.  Their survey of 153 young stellar
objects in the Taurus-Auriga star-forming region revealed a lognormal
distribution of disk masses with a mean mass of $\sim 5 \times 10^{-3}
M_{\sun}$ and a median disk-to-star mass ratio $M_d/M_* = 0.5\%$.  This
result is consistent with \cite{ob95}, who found a median ratio $M_d/M_*
= 0.45\%$ for the 16 classical T-Tauri stars in their sample.  In our
core accretion simulations, the gas surface density decreases as
$\sigma_{gas} \propto 1 / t$.  Our simulations begin at $t = 10^5$
years, so after $10^6$ years, the gas disk mass has decreased by a
factor of 10.  Since the disks in Taurus-Auriga are between 1 and 2 Myr
old (Kenyon \& Hartmann 1995), we assume that the disks studied by
\cite{diskmass} began their lives with 10 times more mass than they
are currently observed to possess.  Therefore, the fiducial point of our
disk mass distribution is $M_d = 0.05 M_{\sun}$.

\cite{diskmass} found that the dispersion in disk mass is 0.5 dex.
However, some of this scatter must be accounted for by the large range of
protostar ages in their sample: According to \cite{kh95}, stars have
been forming at a constant rate in Taurus-Auriga for the last 1-2 Myr.
Work by \cite{shu90} and \cite{lr96} suggests that disks with $M_d > 0.3
M_*$ are gravitationally unstable to non-axisymmetric disturbances, and
will rapidly evolve via accretion to lower-mass configurations.  Even a
disk of mass $0.1 M_*$ can experience low-level gravitational
instability, leading to clumping and fragmentation, if the cooling time
is on the order of local dynamical time anywhere in the disk (Rice,
Lodato \& Armitage 2005, Boss 2002).  In a lognormal distribution with
median $0.05 M_{\sun}$ and standard deviation 0.5 dex, $27\%$ of disks
have initial mass $M_d > 0.1 M_*$.  These disks may be likely to
fragment and form binary star systems.  After fragmentation, planet
formation by core accretion could proceed, but the resulting
circumbinary disk would be substantially reduced in mass, as the
original disk would have donated much of its mass to the newly formed
red or brown dwarf.

In addition, a large dispersion in disk mass raises the likelihood that
even a low-metallicity disk will form a Jovian planet, because the disk
has a high chance of being massive enough to compensate for a low metal
abundance.  Running the simulation with a disk mass dispersion of 0.5 dex
produces a set of synthetic star-disk systems with no apparent relation
between [Fe/H] and likelihood of planet formation.  Since the
planet-metallicity correlation has been well established by
observations, we argue that the spread in initial disk masses around FGK
dwarfs, at the beginning of the T-Tauri phase, is probably not as large
as 0.5 dex.  We select a standard deviation of 0.25 dex for our disk
mass distribution, which means that the violent, global gravitational
instability of $0.3 M_*$ disks is a $3 \sigma$ event.

The core accretion simulations presented in \S \ref{ssd} do not involve
planetary migration, and hence do not assume any particular surface
density profile in the protostellar disk: The planet could be forming in
a region of the disk with locally enhanced or depleted surface density.
To convert from disk mass to surface density (solid+gas), we adopt a
truncated disk with a surface density profile
\begin{equation}
\sigma(r) = \sigma_{in} (r_{in}/r)^{3/2},
\label{sdprofile}
\end{equation}
(Weidenschilling 1977).  For the surface density profile to integrate to
a disk of mass $M_d$, the surface density at the inner boundary must be
\begin{equation}
\sigma_{in} = {M_d / 4 \pi r_{in}^2 \over (r_d / r_{in}) - 1} .
\label{sigmain}
\end{equation}

To find reasonable values for the inner and outer disk radii, we look
first to the mass distribution in the Solar system.  Since the mass of
the terrestrial planets is less than $1\%$ of the mass of the giant
planets, one possible disk configuration can calculated by assuming the
entire disk mass is contained in a region just encompassing the orbits
of Jupiter and Neptune, with $r_{in}$ = 4.5 AU and $r_{out}$ = 36 AU.
We adopt $r_{in} = 4.5$ AU for all synthetic star-disk systems.
Determination of disk outer radii by near-infrared observations, even
from space, is limited by angular resolution, which is not sufficient to
probe inside 100 AU for all but the nearest T-Tauri stars.  The study of
the Trapezium cluster by \cite{trapezium} indicates that $\sim 40\%$ of
disks have radius larger than 50 AU, with a power-law falloff in disk
diameter beyond 50 AU.  However, observations of TW Hydrae (Weinberger
et al. 2002) and GM Aurigae (Schneider et al.  2003) indicate outer disk
radii of $\sim 150$ AU.  We adopt the flat distribution $36 < r_{out} <
100$ AU for outer disk radii in our simulation.

Once the disk mass and outer radius are specified, we use equation
\ref{sdprofile} to calculate the surface density (gas and solid) at 5.2
AU from the protostar.

\subsubsection{Disk Lifetime}
\label{lifetime}

There are many distinct processes that contribute to the
disk dissipation rate in the classical T-Tauri phase.  The gas
photoevaporation rate depends not only on the protostar's accretion
luminosity, but on the radiation environment of the entire star-forming
region.  Even if dust coagulates quickly enough to form planetesimals,
disk gas in dense clusters may be dispersed before runaway accretion
can begin (Hollenbach \& Adams 2004).  Nearby, massive stars
likely accelerate disk dissipation as well (Johnstone et al. 1998).
Photoevaporating winds may also carry micron-size dust grains away
from the disk before they can build larger grains, cutting off
protoplanetary core mass below the threshold of runaway accretion.

We therefore assume that protostellar disk lifetime $T$ follows a
lognormal distribution.  This is consistent with the observations of
\cite{hll01b}, who used $JHKL$ color-color diagrams to derive the disk
frequency in open clusters of different ages, finding that disk
frequency decreases rapidly with cluster age.  By combining their data
with previous observations of open clusters (Haisch, Lada \& Lada 2000,
Kenyon \& Hartmann 1995, and Kenyon \& G\'{o}mez 2001), \cite{hll01b}
found that half of disks are lost by an age of 3 Myr, and almost all
disks dissipate by 6 Myr.  We therefore set the fiducial disk lifetime
at 3 Myr.  In specifying the standard deviation of the disk mass
distribution, we must account for the longevity of disks such as those
in the TW Hydra association, of age 5-15 Myr (Weintraub et al. 2000), or
$\eta$ Cha cluster, with accreting disks up to 10 Myr old (Lawson, Lyo
\& Feigelson 2003).  These disks are observationally rare, but this may
be a selection effect, because the detection of pre-main-sequence
objects is biased toward young, massive disks (Lawson, Lyo \& Feigelson
2003).  We therefore set the spread of the disk lifetime distribution at
0.15 dex, which places 6 Myr disks at $2 \sigma$ above the fiducial, and
the 10 Myr disks of $\eta$ Cha at $3.5 \sigma$ above the fiducial disk
lifetime.


\subsubsection{Chemical Composition and Calculation of Solid Surface
Density}
\label{comp}

The chemical composition of our star-disk systems was specified by
[Fe/H] and [Si/Fe], which we assume to be equal to [$\alpha$/Fe].
These, in conjunction with the chemical model of HGL and equation
\ref{sdprofile}, specify the solid surface density at Jupiter's
distance from the Sun.  We consider only elements with sufficient
abundance in the sun (according to Anders \& Grevesse 1989) that $N_{\rm
X} \geq 10^{-6} N_{\rm H}$, where $N_{\rm X}$ is the number of atoms of
a element X present in one cubic centimeter of disk material, and $\log
\, N_{\rm H,\sun} = 12.00$.  The elements in our simplified star-disk
systems are therefore H, He, C, N, O, Ne, Na, Mg, Al, Si, S, Ar, Ca, Fe
and Ni, of which He, C, O, Ne, Mg, Si, S, Ar and Ca are
$\alpha$-elements or related products of helium burning.

\begin{figure}
\epsscale{0.6}
\plotone{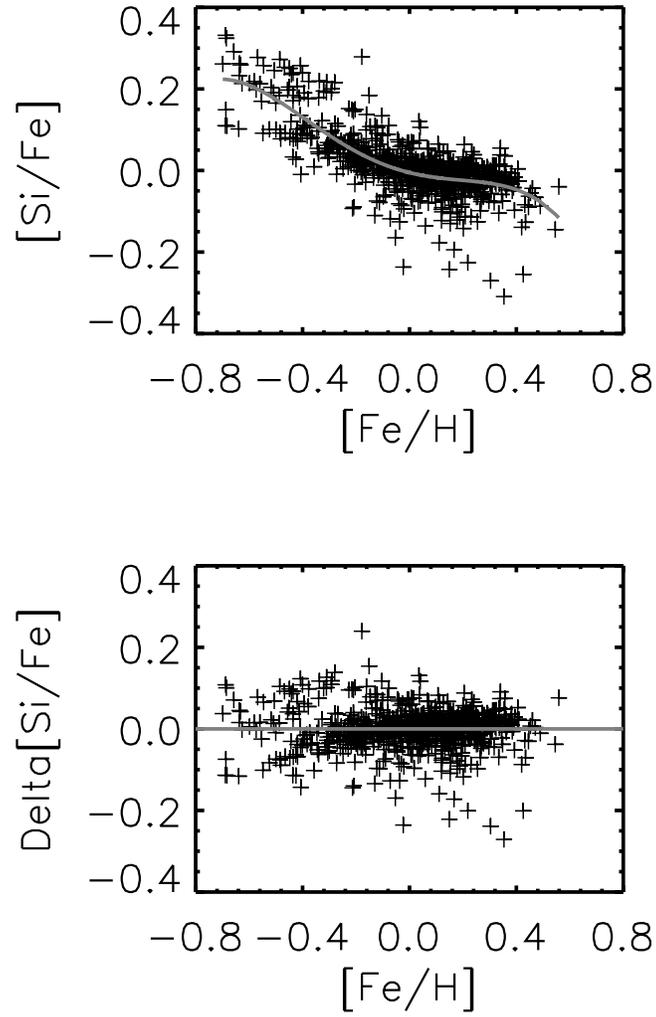}
\caption{Top: Empirical model of [Si/Fe] as a function of [Fe/H].
+ signs represent the SPOCS data; the gray, solid line is a fitted
fourth-order polynomial.  Bottom: Fit residuals.  There is a slight
trend toward increasing $\Delta {\rm [Si/Fe]}$ where ${\rm [Fe/H]} <
-0.3$, but for the most part the fit residuals are independent of
[Fe/H].}
\label{poly_resid}
\end{figure}

\begin{figure}
\plotone{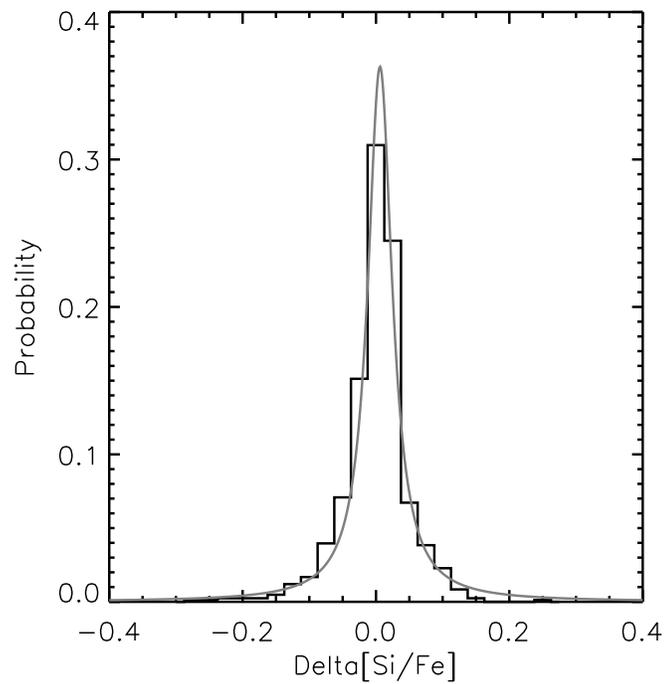}
\caption{Model of residuals around fiducial [Si/Fe]([Fe/H]).  $\Delta
{\rm [Si/Fe]}$ is best characterized by a Cauchy distribution, which is
used in our simulation to randomly select the deviation in [$\alpha$/Fe]
from the fiducial for each star-disk system.}
\label{cauchy}
\end{figure}

$\alpha$-abundances for our star-disk systems cannot be selected
independently of [Fe/H], since Galactic chemical evolution assures that
[$\alpha$/Fe] will depend on [Fe/H] (Timmes, Woosley \& Weaver 1995).
To use [$\alpha$/Fe] as a random variable in our simulation, we use the
\cite{FV05} subset of SPOCS with a uniform probability of planet
detection to empirically model [Si/Fe] vs. [Fe/H].  We first fit a
fourth-order polynomial to [Si/Fe] as a function of [Fe/H] in the
uniform set:
\begin{equation}
{\rm [Si/Fe]} = -0.00516 - 0.162 {\rm [Fe/H]} + 0.449 {\rm
[Fe/H]}^2 - 0.377 {\rm [Fe/H]}^3 - 0.974 {\rm [Fe/H]}^4 .
\label{poly4}
\end{equation}
We subtract this fit to find the iron abundance-independent [Si/Fe]
residuals $\Delta {\rm [Si/Fe]}$.  The polynomial fit and residuals are
plotted in Figure \ref{poly_resid}.  We then fit a Cauchy distribution
to a histogram of $\Delta {\rm [Si/Fe]}$:
\begin{equation}
P(\Delta {\rm [Si/Fe]}) = C
\left ( {(1/2) \Gamma \over (\Delta {\rm [Si/Fe]} - (\Delta {\rm
[Si/Fe]})_0)^2 + ((1/2) \Gamma)^2} \right ),
\label{cauchyeq}
\end{equation}
where $C = 0.0254$, $(\Delta {\rm [Si/Fe]}_0) = 0.00600$ is the center
of the distribution, and $\Gamma = 0.0445$.  This fit is shown in Figure
\ref{cauchy}.  The fact that the center of the residual distribution is
so close to zero means the fiducial sequence [Si/Fe] as a function of
[Fe/H] is well modeled.  Since we assume ${\rm [\alpha/Fe] = [Si/Fe]}$,
the Cauchy distribution describing $\Delta {\rm [Si/Fe]}$ is the
distribution from which the random variable describing
$\alpha$-abundance is drawn.

The synthetic star-disk systems are initially given Solar abundances.
Then iron abundance of each synthetic star-disk system is selected from
a uniform distribution within the limits $-0.7 \leq {\rm [Fe/H]} \leq
0.5$.  All known extrasolar planets orbit stars within this range.  The
abundances of all metals are then altered by a factor of [Fe/H]:
\begin{equation}
N_{\rm M} = 10^{\rm [Fe/H]} N_{\rm M,\sun} ,
\label{fehfactor}
\end{equation}
where $N_{\rm M}$ denotes metal abundance.  Using equation \ref{poly4},
the fiducial value of [$\alpha$/Fe] for the selected [Fe/H] is
calculated.  From the distribution in equation \ref{cauchyeq}, we then
select the $\alpha$-adjustment factor $\Delta {\rm [\alpha/Fe]}$.  We
are then ready to specify [$\alpha$/Fe] for the star-disk system:
\begin{equation}
[\alpha/{\rm Fe}] = [\alpha/{\rm Fe}]_{\rm fiducial} + \Delta {\rm
[alpha/Fe]}.
\label{afesum}
\end{equation}
The individual $\alpha$-element abundances for the star-disk system are
then altered by a factor of [$\alpha$/Fe]:
\begin{equation}
N_{\alpha} = 10^{[\alpha/{\rm Fe}]} N_{\alpha,0} .
\label{alphafactor}
\end{equation}

Once the abundances of all the elements in the star-disk system are
specified, we can calculate the gas/solid ratio at the snow line.  As in
\S \ref{ssd}, the solid elements are Na, Mg, Al, Si, S, Ca, Fe, Ni,
62\% of oxygen and the hydrogen bound up in H$_2$O.  The gas/solid ratio
is calculated as
\begin{equation}
G/S = { \sum_X (1-S_X) W_X N_X \over \sum_X S_X W_X N_X},
\label{gassolid}
\end{equation}
where $W_X$ is the atomic weight of element X, $S_X$ is the
fraction of X that is solid at the ice line, and $N_X$ is the abundance
of X.  Knowing the gas/solid ratio and the total surface density at 5.2
AU calculated from equation \ref{sdprofile}, we can then calculate the
solid surface density:
\begin{equation}
\sigma_{\rm solid} = {\sigma \over (G/S + 1)}.
\label{sigmasolid}
\end{equation}
As one example of the relationship between [Si/Fe] and $\sigma_{\rm
solid}$ in our simulation, a disk with [Fe/H] = 0.0 and [Si/Fe] = 0.3
has solid surface density $37\%$ higher than a disk with the same mass
and outer radius, with [Fe/H] = 0.0 and [Si/Fe] = 0.0.  This is less
than the solid surface density obtained by increasing only oxygen
abundance because helium and carbon, gaseous $\alpha$-elements, have
their abundances increased along with silicon and oxygen.

At this point, we have specified the solid surface density and the
lifetime of our synthetic star-disk system.  Using equation
\ref{formtimeeq}, we calculate the timescale for planet formation.  If
this timescale is less than the protostellar disk lifetime, a planet
forms.  By building $10^7$ synthetic star-disk systems and analyzing
their properties, we now perform a theoretical test of the
planet-silicon correlation and compare the results with the trends
observed in the SPOCS data.

\subsection{Simulation Results}
\label{modelresults}

In this section, we present the results of our Monte Carlo simulations.
We assess how well the simulations reproduce the observed
planet-metallicity correlation.  We then use the disk lifetime and time
to rapid gas accretion from our simulations to predict the fraction of
Jovian planets that do not migrate, and therefore may dominate planetary
systems similar to our own.  Next, we estimate the fraction of Saturn
analogs at 5 AU from their parent star, and use our relation between
solid surface density and time to rapid gas accretion to calculate the
lifetime of the solar nebula.  Finally, we examine the planet-silicon
correlation as predicted by our simulations and speculate on other
elements that might correlate with likelihood of planet formation.

\subsubsection{Planet-Metallicity Correlation}

The first test of our simulation is whether it reproduces the
planet-metallicity correlation.  This correlation has been well
established by many observers (e.g Fischer \& Valenti 2005), and has
also been reproduced by other semi-analytic planet formation models in
the literature, including those of Ida and Lin (2004a, 2004b, 2005) and
\cite{kornet05}.  Both the Ida \& Lin and the Kornet et al. theories
incorporate results from core accretion calculations that are very
similar to those used in our simulation.  In particular, all three
theories contain a phase of rapid gas accretion, whose time of onset
depends sensitively on the solid surface density.  Hence, for a given
disk mass and lifetime, these theories naturally produce a connection
between host-star metallicity and the presence of a detectable
Jovian-mass planet.  The Ida-Lin theory also explicitly reproduces the
observed paucity of planets with masses that fall in the $100-200
M_{\earth}$ range, in which rapid gas accretion is expected to occur
most easily.

\begin{figure}
\plotone{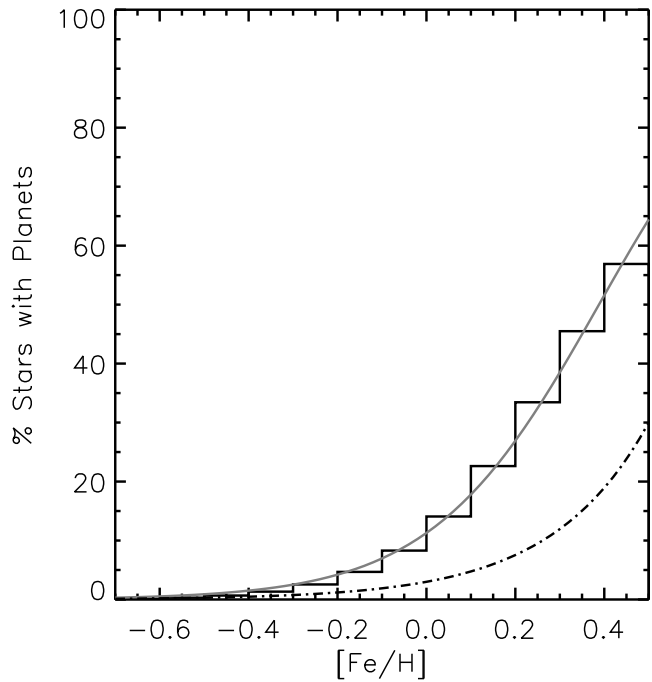}
\caption{Planet-metallicity correlation as predicted by Monte Carlo
simulations.  The stair-stepped black line is a histogram showing the
percent of stars with planets in each metallicity bin, as predicted by
our simulations.  Overplotted in smooth gray is our fit of a logistic
function to the analytical planet-metallicity correlation.  The dotted
line is the planet-metallicity correlation measured by \cite{FV05}. }
\label{pmc}
\end{figure}

We calculate the fraction of stars in each metallicity bin that have
planets, and compare with equation \ref{pmceq}, derived by \cite{FV05}.
Figure \ref{pmc} shows that our simulation reproduces the increasing
trend in number of planets with [Fe/H].  However, above super-Solar
metallicity, our simulation shows the slope of the planet-metallicity
correlation changing from increasing with [Fe/H] to decreasing with
[Fe/H].  The planet-metallicity correlation takes the form of a logistic
curve, as expected for a monotonically increasing relation bounded by
$0\%$ (no stars form planets) and $100\%$ (all stars form planets).
Indeed, it appears that above [Fe/H] = 0.3, the slope of our theoretical
planet-metallicity correlation starts to flatten.  We determine the
following equation for the probability that the disk surrounding a star
of metallicity [Fe/H] ever had a giant planet companion:
\begin{equation}
\label{anpmc}
P({\rm planet}) = {1 \over 7.86 (0.00493^{\rm [Fe/H]}) + 1.00}
\end{equation}

The inflection point of equation \ref{anpmc} is at ${\rm [Fe/H]} = 0.38$
dex.  At [Fe/H] = 0.38 dex, fiducial [$\alpha$/Fe] = -0.04, and given an
accretion disk with our simulation's fiducial disk mass ($M_d = 0.05
M_{\sun}$), and the median outer radius in our simulation (68 AU), the
solid surface density at Jupiter's distance from the sun is $\sigma_{\rm
solid} = 11.1$\ g cm$^{-2}$.  This is on the section of the planet
formation timescale curve in Figure \ref{formtime} where $t_{\rm rga} <
3$ Myr, where a large increase in $\sigma_{\rm solid}$ lowers the
formation time only slightly.  Above [Fe/H] = 0.38 dex, therefore,
increasing metallicity, despite the fact that it increases solid surface
density, can longer profoundly increase the probability of planet
formation.

\subsubsection{Long-Period vs. Short-Period Planets}

Our simulations predict that $16\%$ of FGK dwarfs have giant planets, in
rough agreement with \cite{marcy05}, who conclude that $\sim 12\%$ of
FGK dwarfs should have planets.  \cite{marcy05} used a flat
extrapolation between 5 and 20 AU of the observed distribution of planet
frequency vs. semimajor axis to predict the frequency of giant planets
around FGK dwarfs.  This distribution of planet frequency as a function
of semimajor axis in the Lick/Keck/AAT planet search data is not well
constrained.  Only 47 of the planet hosts in these data had at at least
10 observations, with precision 30 m s$^{-1}$, spanning four years.  The
decline in planet frequency above semimajor axes of 3 AU may thus be an
artifact of the finite time baseline of the observations.
\cite{marcy05} report increasing incompleteness of the data beyond 3 AU.


We can calculate the difference between protostellar disk lifetime and
time to rapid gas accretion, $L - t_{rga}$, and compare this quantity
with Type II migration timescales in the literature.  This way, we can
use our simulation results to estimate of the fraction of Jupiter-mass
planets that stay approximately at their formation radius, and do not
migrate.  Estimating the number of Jovian-mass planets that do not
migrate gives an upper limit to the number of planetary systems that
might be Solar System analogs.  It also has implications for the yield
of long time-baseline Doppler surveys.  For Jovian planets to be
detected by radial-velocity searches, the gas in the disk had to remain
long enough after the planet reached approximately Jovian mass to force
the planet to migrate inward.  \cite{nelson00}, \cite{dangelo03} and
\cite{pn05} are all in agreement that the migration timescale of a
Jupiter-mass planet at 5.2 AU from the sun is $10^4$ orbital periods, or
$\sim 10^5$ years.  The long-period planets in our simulation, which
never experience Type II migration, are those in which the planet
reached rapid gas accretion less than $10^5$ years before the disk
dissipated.  These account for $4\%$ of the planets formed in our Monte
Carlo simulations.  True Solar-System analogs, therefore, likely make up
only a small percentage of the planetary systems that initially formed
in the Solar neighborhood, though they may account for a higher
percentage of planetary systems that survived to maturity, since some
giant planets end Type II migration by colliding with their parent stars
(Ida \& Lin 2004b).

\begin{figure}
\plotone{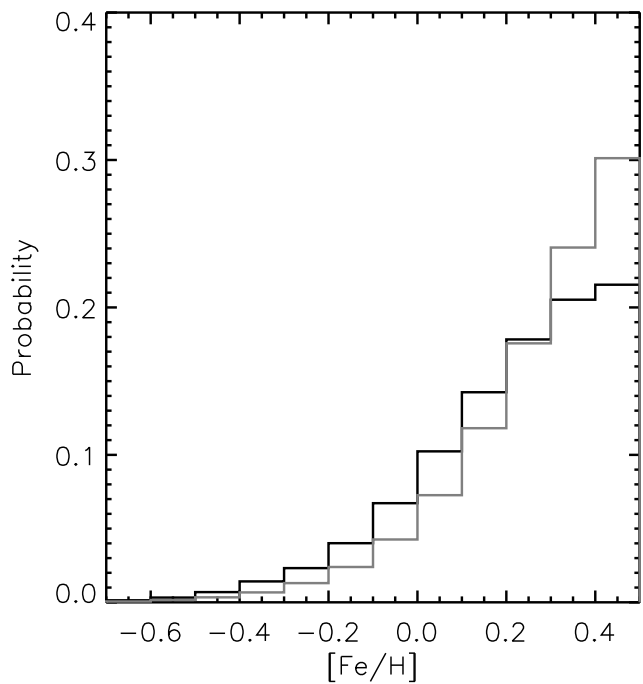}
\caption{[Fe/H] distribution of hosts of long-period (black) and
short-period (gray) planets in our simulation results.  Hosts of
short-period planets tend to be more metal-rich than hosts of
long-period planets.}
\label{period}
\end{figure}

Even as we predict a higher frequency of planets than has been observed,
as expected from lack of completeness for long periods in Doppler
surveys, when we compare our theoretical planet-metallicity correlation
with the observations of \cite{FV05}, we find that the discrepancy
between our simulation and observations is highest at low metallicities
(Figure \ref{pmc}).  At [Fe/H] = -0.2, we predict the frequency of giant
planet formation is 3.5 times what is observed, whereas at [Fe/H] = 0.4,
we predict this frequency is only 2.7 times what is observed.
Low-metallicity disks that manage to form planets are more likely than
their high-metallicity counterparts to dissipate soon after the planets
have reached rapid gas accretion, due to the overall longer timescale of
planet formation in these disks.  This would explain our simulations'
prediction of a higher number of planets than observed being more
pronounced at low metallicities.  We assume planets that finish forming
less than the Type II migration timescale before disk dissipation, and
thus do not migrate, are long-period planets, while the rest are
short-period planets.  Metal-poor disks, therefore, have a higher
probability of forming as yet undetected long-period planets than
metal-rich disks.  Figure \ref{period} compares the [Fe/H] distribution
of the host stars of long-period and short-period planets in our
simulation re3sults.  The [Fe/H] distribution of short-period planets is
more strongly skewed toward high metallicities than that of long-period
planets.


\subsubsection{Saturn-Mass Planets at 5 AU}

The combination of the core accretion process of planet formation and
rapid dissipation of gas in protostellar disks would allow for the
existence of planets like Saturn, which has a mass consistent with
having suddenly stopped gathering an envelope in the middle of its rapid
gas accretion phase.  One formation scenario for Jupiter that is
consistent with core accretion theory is that Jupiter was embedded in a
disk in which a substantial amount of gas remained for 5.2 Myr, then
suddenly dissipated before the planet could experience Type II
migration; Jovian planets are then the result of a coincidence between
the gas dissipation epoch and the rapid gas accretion stage of planet
formation.  (If Jupiter and Saturn were nearer each other than their 2:1
mean motion resonance, this would also prevent substantial inward
migration; see Morbidelli, Crida \& Masset [2005].)  The
Jupiter-migration timescale of $10^5$ years would then also be the upper
limit on gas dissipation time in a rapid-dissipation scenario.
\cite{pn05} calculate that once a planet has reached the rapid gas
accretion phase (having a typical mass of $\sim 30 M_{\earth}$, in
agreement with LBA), a Jupiter mass of gas can be accreted in $10^3$
years, for $1\%$ interstellar dust opacity.  Saturn-mass planets, if
they indeed are the result of gas dissipation occurring during the {\it
middle} of rapid gas accretion, should then account for $10^3 / 10^5 =
1\%$ of the population of giant planets on ten-year orbits.  This
speculation, of course, does not take into account the effects of two
embryos competing for gas, which could also suddenly cut off rapid gas
accretion.

\subsubsection{Nature of the Solar Nebula}

\begin{figure}
\plottwo{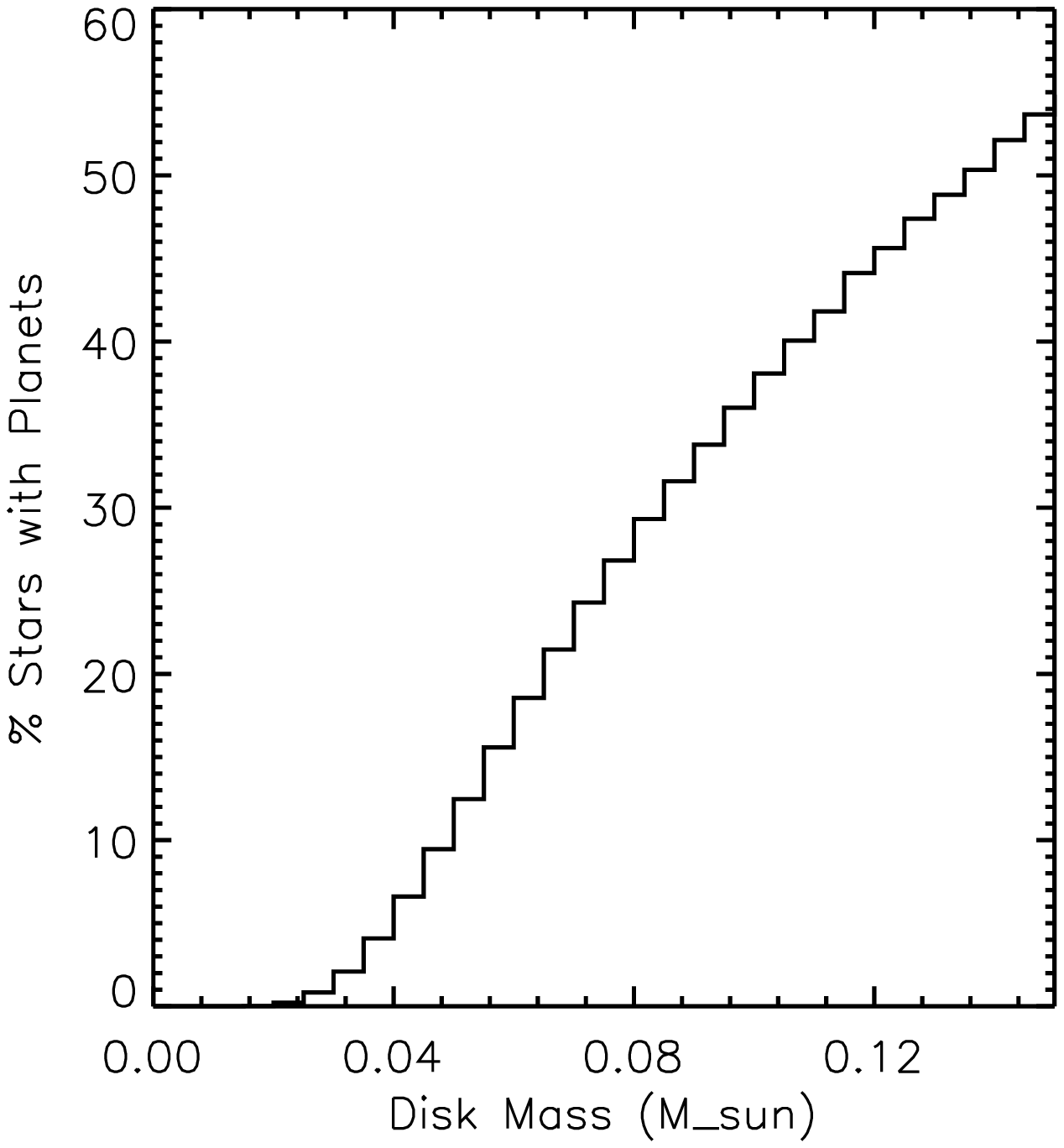}{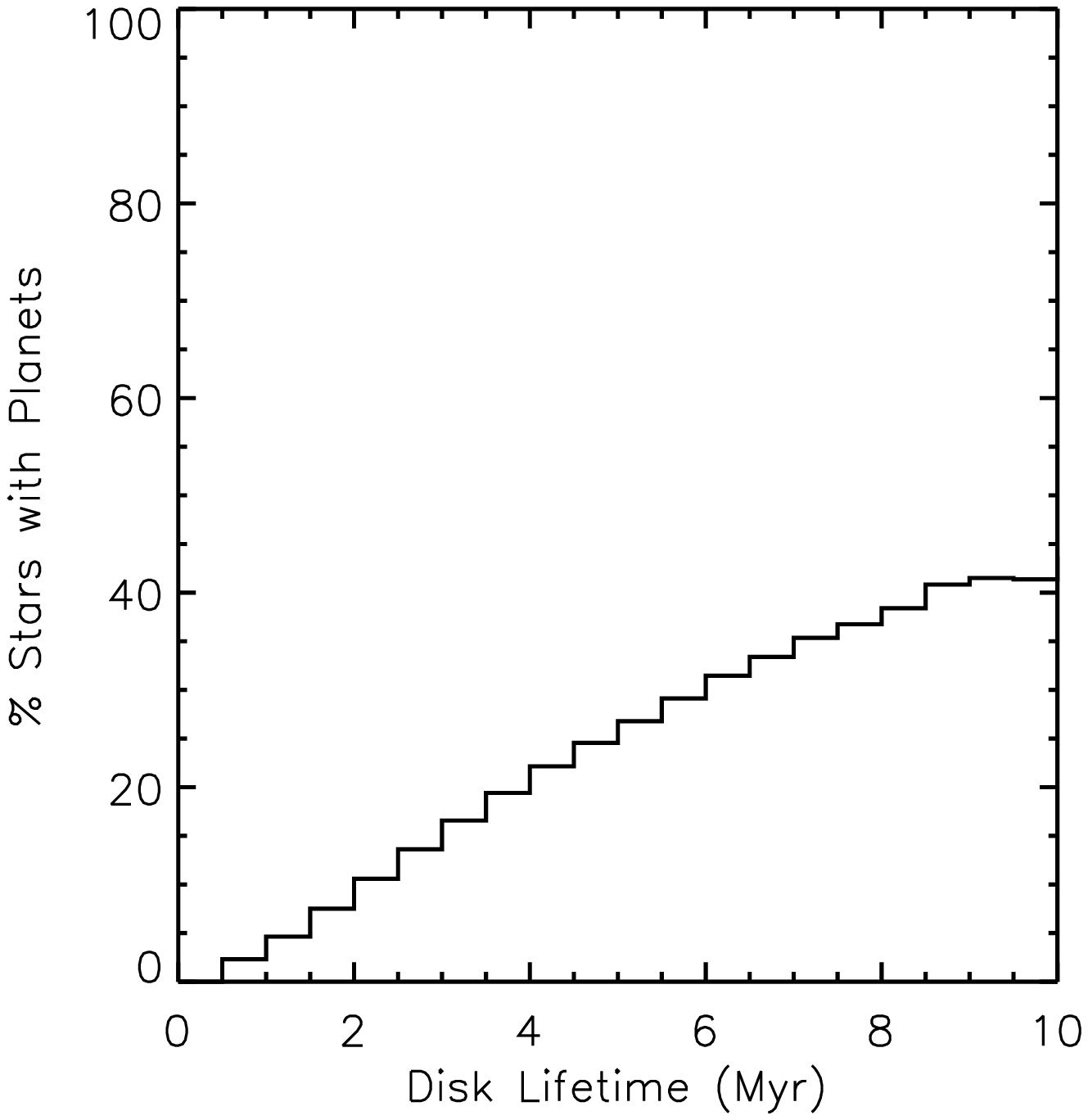}
\caption{Left: Percent of stars with planets as a function of initial
disk mass.  Right: Percent of stars with planets as a function of disk
lifetime.  Our core accretion simulations indicate that the Solar nebula
lasted for 5.2 Myr, at which $27\%$ of stars form planets.}
\label{masslife}
\end{figure}

We now use our numerical simulations of the relationship between
$\sigma_{\rm solid}$ and $t_{\rm rga}$ to speculate on the nature of the
Solar nebula.  Saumon and Guillot (2004) conclude that the upper limit
on solid surface density during Jupiter's formation should be 8 g
cm$^{-2}$, based on the abundances of heavy elements in Jupiter's
envelope.  Based on our core accretion simulations, this corresponds to
a time to rapid gas accretion of 5.2 Myr.  Our simulations therefore
predicts that the lifetime of the Solar nebula was $1.6\sigma$ above
that of the average Population I disk.  The Sun's formation environment,
therefore, may have been somewhat more quiescent than typical.  Assuming
the distribution of mass in the Solar nebula roughly matches that of the
Solar system today, the inner and outer disk radii during Jupiter's
formation were at 4.5 and 36 AU, just encompassing the orbits of Jupiter
and Neptune.  This gives the total mass of the Solar nebula as $0.05
M_{\sun}$.  Figure \ref{masslife} gives the percent of stars that form
Jovian-mass planets as a function of disk mass $M_d$ and lifetime $T$.
According to our Monte Carlo simulations, $27\%$ of disks with the
lifetime of the Solar nebula will form planets, whereas only $10\%$ of
disks with the mass of the Solar nebula will form planets.  Since the
Sun is neither particularly metal-rich, in comparison to the population
of planet hosts, nor $\alpha$-rich, the critical factor that allowed
planets to form in the Solar nebula was its long lifetime.

\subsubsection{Planet-Silicon Correlation}

\begin{figure}
\epsscale{0.6}
\plotone{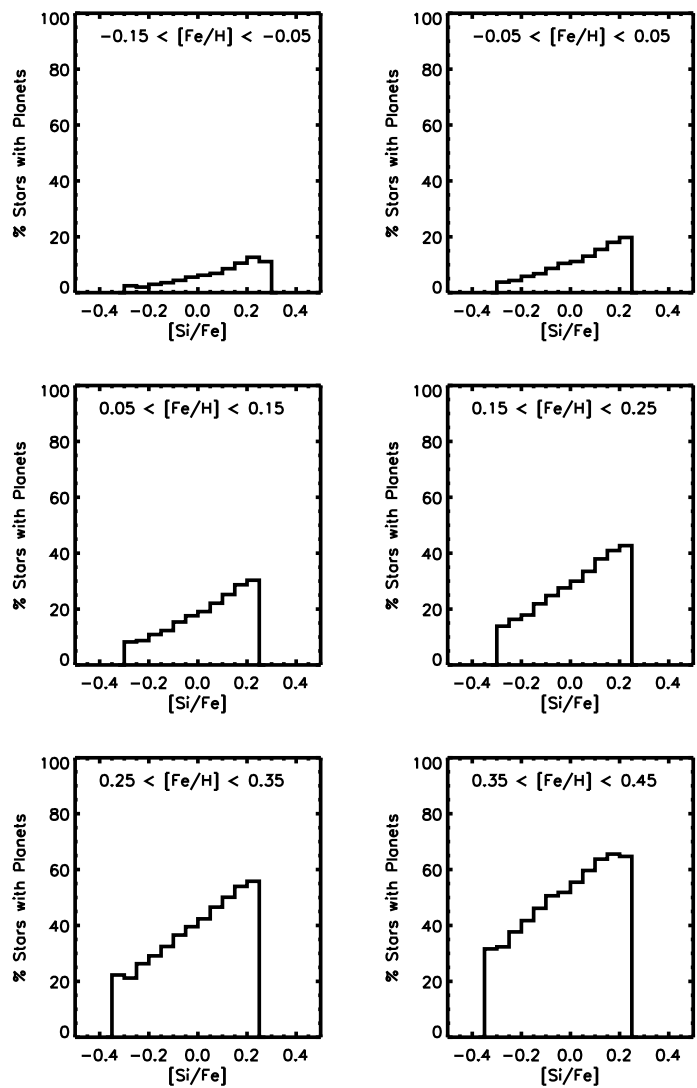}
\caption{Percent of stars with planets as a function of [Si/Fe] at
constant [Fe/H], predicted by simulations.  We model the fraction
of stars with planets is a linear function of [Si/Fe] with a slope
increasing with [Fe/H].}
\label{model_psicorr}
\end{figure}

Most importantly, our simulations predict that stars with planets are
silicon-enhanced.  Figure \ref{model_psicorr} shows histograms of the
fraction of stars with planets as a function of [Si/H], in different
[Fe/H] bins.  Instead of a power law with a large exponent, the
probability of planet formation appears to increase linearly with
[Si/Fe] at a constant value of [Fe/H], or logarithmically with silicon
abundance.  There is some sign that the planet-silicon correlation
assumes a logistic form (as, with finite top and bottom bounds, it
should), especially in the [Fe/H] range $-0.15 < {\rm [Fe/H]} < 0.05$.
The planet-silicon correlation is steeper at higher [Fe/H].  We model
the correlation as the linear function
\begin{equation}
\left [ P({\rm planet}) \right ]_{\rm [Fe/H]} = C + A {\rm [Si/Fe]},
\label{psicslope}
\end{equation}
$P$ is the percent of stars with planets where $C$ is a constant
reflecting the change in the total number of planets expected to form in
each [Fe/H] bin.  For ${\rm [Fe/H]}$ = -0.15, -0.05, 0.05, 0.15, 0.25
and 0.35, the slope $A$ of the planet-silicon correlation is 0.19, 0.32,
0.47, 0.60, 0.66 and 0.67.  The increasing slopes may reflect that the
low-[Fe/H] bins include some of the flat bottom of the logistic curve.

At this point, we would like to assess how well the planet-silicon
correlation predicted by our simulations matches the SPOCS data.  As
discussed in \S \ref{expsim}, the small number of planet hosts in SPOCS
prevent us from merely binning the data by [Fe/H] and measuring the
planet-silicon correlation in each bin, as we did in Figure
\ref{model_psicorr} for our simulation results.  However, the fiducial
value of [Si/Fe] predicted by equation (\ref{poly4}) changes little in
the range $0.05 < {\rm [Fe/H]} < 0.35$: here, [Si/Fe] vs. [Fe/H] is
almost flat.  We can, therefore, measure the planet-silicon correlation
in the SPOCS data using the 447 stars, including 56 planet hosts, in
this [Fe/H] range.

\begin{figure}
\plotone{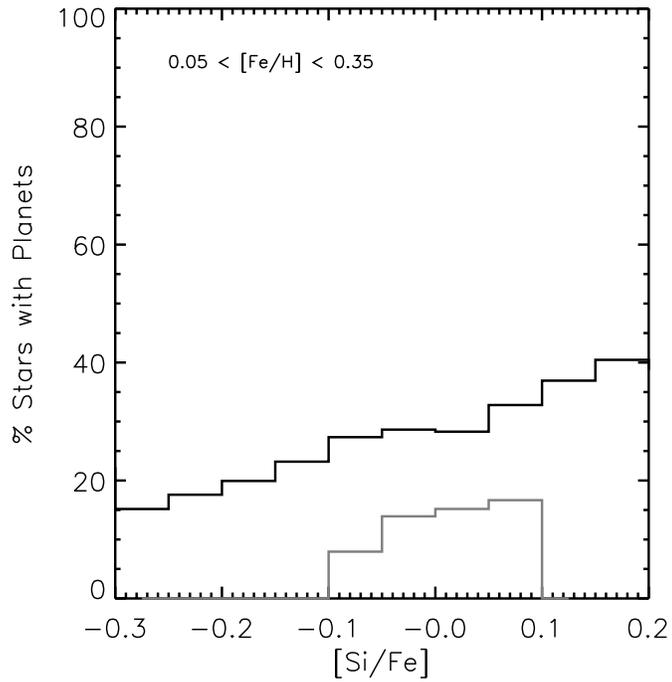}
\caption{Comparison between theoretical (black) and observed (gray)
percent of stars with planets, as a function of [Si/Fe], for $0.05 <
{\rm [Fe/H]} < 0.35$.  For stars in this [Fe/H] range, our
simulations predict that the percent of stars with planets is
a linear function of [Si/Fe] with a slope of 0.51.  In the [Si/Fe] bins
where the number of planets in SPOCS is nonzero, the slope of the
observed planet-silicon correlation is 0.54.}
\label{comparison}
\end{figure}

Figure \ref{comparison} compares the planet-silicon correlation
predicted by our simulations with that present in the SPOCS data for
stars with $0.05 < {\rm [Fe/H]} < 0.35$.  Modeling the SPOCS
planet-silicon correlation with a linear fit {\it over the range of
[Si/Fe] values where planets have been found} ($-0.1 < {\rm [Si/Fe]} <
0.1$) gives the equation
\begin{equation}
P({\rm planet})_{\rm SPOCS} = 0.13 + 0.55 {\rm [Si/Fe]} .
\label{spocs_psicorreq}
\end{equation}
A linear fit to the theoretical planet-silicon correlation implied by
our simulations for all values of [Si/Fe] in the range $0.05 < {\rm
[Fe/H]} < 0.35$ gives the function
\begin{equation}
P({\rm planet}) = 0.29 + 0.51 {\rm [Si/Fe]} .
\label{model_psicorreq}
\end{equation}
Therefore, although we predict a higher frequency of planets than
observed, the slope of the planet-silicon correlation implied by our
simulations reproduce the slope of planet-silicon correlation present in
the SPOCS data, at least for metal-rich stars.

The difference between the empirical fit to the planet-silicon
correlation in the above paragraph and that of \S \ref{expsim} is
whether the non-detections of planets at low values of [Si/Fe] are
assumed to be significant.  Examination of Figure \ref{sifehists} shows
that the lowest $\sim 0.1$ dex of the [Si/Fe] range in each [Fe/H] bin
contains no observed planets.  This influences our process of finding
the power-law exponent that best describes the SPOCS planet-silicon
correlation by making the percent of stars with planets resemble a step
function in [Si/Fe].  As discussed in \S \ref{expsim}, any attempt to
fit a power law to a step function will force the power law to assume a
high exponent.  When we consider only the range of [Si/Fe] values where
planets have been found, the planet-silicon correlation in SPOCS assumes
an approximately linear form with a modest slope.

The paucity of detected planets at low values of [Si/Fe], which in \S
\ref{expsim} forces the observed planet-silicon correlation to appear
steeper than we predict it should be, may, as in the case of the
planet-metallicity correlation, reflect observational bias against
long-period planets.  The Jovian planets that form around $\alpha$-poor
stars are, as in the case of iron-poor stars, most likely to finish
forming too late to experience Type II migration.  These planets are
therefore less likely to be detected than the Jovian planets orbiting
$\alpha$-rich stars.


\subsubsection{Other Elements that may Correlate with Planet Frequency}

To the extent that titanium can be said to share properties of the pure
$\alpha$-elements, the question of why planet hosts do not appear to be
titanium enhanced is puzzling.  If [O/Fe] does actually decline with
increasing [Fe/H] in the range $-0.8 < {\rm [Fe/H]} < 0.5$, then
[Ti/Fe], which exhibits the same declining trend in the SPOCS data,
might actually be the best tracer of [O/Fe]---though other observers,
including \cite{abtrends} and \cite{bodaghee03} have found that [Ti/Fe]
as a function of [Fe/H] is flat for metal-rich stars.  Since the assumed
correlation between silicon and oxygen abundance is the basis for our
prediction that silicon enhancement in planet hosts would result from
planet formation by core accretion, we would expect planet hosts to be
titanium-enhanced if [Ti/Fe] is the best tracer of [O/Fe].  Silicon,
however, may not derive its predictive power of planet formation
exclusively from a correlation with oxygen.  At Jupiter's position in a
planet-forming disk of Solar composition, silicon is the third most
abundant solid material at the ice line, after oxygen and iron.  Even if
a disk has low [O/Fe], a high value of [Si/Fe] may give it enough solid
material to form a planet.  This is a property that titanium cannot
share, as it is $\sim 400$ times less abundant than silicon in
solar-type stars.

Since we are assuming that [Si/Fe] is a proxy for [$\alpha$/Fe], one
test of our simulation results would be to replicate the analysis in \S
\ref{observed} on other $\alpha$-elements.  We predict that planet hosts
have enhanced abundances of C, Ne, Mg, S, Ar, S, and most especially O.
We also suspect that enrichment is most likely to be observed in the
subset of the $\alpha$-elements that, like silicon, are particularly
important grain-forming materials: oxygen, carbon and magnesium.  For
example, as a noble gas, argon can have no direct influence on planet
formation and would function only as an indicator of the presence of
other core-forming materials.  Though the fiducial [Ar/Fe] vs. [Fe/H]
sequence follows the same general pattern as [O/Fe] vs. [Fe/H], there is
scatter in both these relations, so the Ar/O ratio changes somewhat
among metal-rich stars.  This scatter would lessen the utility of argon
abundance as a predictor of planet formation.  The diluting effect of
scatter in Si/O on the planet-silicon correlation could be counteracted
by the fact that silicon is an important core-forming material in its
own right.  Since our simulation assumes that Si/O is constant in all
metal-rich stars, our prediction is that the planet-oxygen correlation
has the same slope as the planet-silicon correlation, such that
\begin{equation}
P({\rm planet})_{\rm model} = C + 0.51 {\rm [O/Fe]} .
\label{oprediction}
\end{equation}

Our simulation does not explore the effects of enhancing
abundant non-$\alpha$ elements such as nickel.  In the simplest
interpretation of the core-accretion theory, where all solids are
equally useful for forming planets, the planet-nickel correlation should
be much weaker than the planet-silicon correlation, because of the low
abundance of nickel compared with silicon and oxygen.  Our statistical
analysis indicates that [Ni/Fe] values are statistically enhanced for
planet-bearing stars by a degree similar to that observed for silicon.
If this result persists when more stars have been observed, it may
indicate that not all solids are equally good for planet formation: some
may accelerate or retard core growth through microchemical processes.


\section{Conclusion}

We find that there is statistical evidence for silicon and nickel
enrichment of planet hosts in the SPOCS data of \cite{SPOCS}.  Although
the planet hosts do not exhibit any anomalous abundance patterns that
indicate that they might be members of a different stellar population
than other field stars, they do have values of [Si/Fe] and [Ni/Fe] that
are systematically enhanced over other stars of the same [Fe/H].

We have constructed Monte Carlo simulations that predict the likelihood
of forming planets by core accretion in disks of differing mass,
lifetime, outer radius and chemical composition.  Our simulations
reproduce both the planet-metallicity correlation and the planet-silicon
correlation reported in this work.  The simulation demonstrates that the
abundance patterns of planet hosts in the SPOCS data are consistent with
planets having formed by core accretion.  According to the core
accretion theory, planets form most easily in ice-rich disks.  We
predict that the observed silicon enhancement of planet hosts arises
primarily from the correlation in the abundances of silicon and oxygen,
and secondarily from silicon's own importance as a grain-forming
material.  The planet-silicon correlation would naturally arise if the
core of Jupiter is made mainly of water ice frozen on silicate and/or
iron grains.  (See, however, \cite{tar}, which argues that Jupiter's
core may have formed from a tar-like mixture of organic compounds.)

More observations are needed to determine whether the planet-silicon
correlation is robust, and whether it truly takes the form predicted by
our simulations.  Despite its large size, the SPOCS catalog is seriously
affected by small-sample statistics when the planet hosts are separated
into [Fe/H] bins to have their abundance patterns analyzed.  The
tantalizing suggestion that a silicon correlation may exist in the data
underscores the need for a larger pool of stars from which to draw
planet hosts and control sets. Our simulation predicts that the silicon
effect should persist in a larger sample.  Although we do not directly
simulate variations in nickel abundance, the idea that nickel
enhancement increases likelihood of planet formation is consistent with
the core accretion theory.  However, if all solids form protoplanetary
cores equally well, the planet-silicon correlation should be much
stronger than the planet-nickel correlation.


The most immediate test of our simulation is whether planet hosts show
oxygen enhancement.  Our simulation relies on the assumed correlation
between oxygen and silicon abundances, so our analytical framework would
be invalidated if a SPOCS-type survey shows no evidence for oxygen
enrichment in planet hosts.  Other $\alpha$-elements should also be
correlated with the presence of planets, particularly the important
core-forming materials C and Mg.

The question of what fraction of Pop I stars have at least one giant
planet has been much debated recently.  Our prediction of
planetary systems around $16\%$ of FGK dwarfs roughly agrees with the
\cite{marcy05} extrapolation of the planet frequency-semimajor axis
relation in the Lick/Keck/AAT planet search data.  According to our
calculation of the fraction of giant planets that experience Type II
migration, planets on Jupiter-like orbits account for at least $4 \%$ of
giant planet systems.  A Solar system-like configuration of planets,
while not particularly common, is nevertheless a viable outcome of the
core accretion-migration scenario of planet formation.   Detection of
planets by transit searches and direct imaging will aid in solving the
problem of completeness in discoveries of planets at 3-20 AU orbits,
since these survey methods are sensitive to planets in a complementary
part of the mass and semimajor axis parameter space to current Doppler
surveys.

This work was supported by the National Science Foundation Graduate
Research Fellowship awarded to S. R., by the NASA Origins of Solar
Systems program grant NNG04GN30G and a National Science Foundation
Career Grant to G. L., and by NASA grant NAG-5-13285 and NSF grant
AST-0507424 to P. B.

\appendix

\section{Statistical Methods: Determining Significance of Planet-[Si/Fe]
Correlation}
\label{appendix1}

We used the following procedure to find $D_{\rm X}$\ and $D_{\rm Fe}$:
\renewcommand{\labelenumii}{\alph{enumii}}
\begin{enumerate}

\item Generate a bootstrap realization of SPOCS, a compilation of
stellar abundances randomly selected from SPOCS.  Choose stars at random
from the SPOCS catalog to form set $A$, which will have the same number
of members as the SPOCS catalog.  Each star is selected independently of
previously selected members of $A$, thus allowing duplicate entries in
$A$.  $A$\ will include both planet hosts and planetless stars
in roughly the same proportions as in the SPOCS catalog.  The set $A$ is
therefore a synthetic SPOCS compilation, created by sampling the SPOCS
catalog with replacement.

\item Identify the planet hosts in $A$.  These will comprise set $B$.
The number of occurrences of each planet host in $A$ is preserved in
$B$, so where the same SPOCS planet host is present more than once in
$A$, it will also be duplicated in $B$.

\item Find the iron-metallicity distribution of the planet hosts in the
bootstrap realization of SPOCS by creating a normalized histogram of the
[Fe/H] values of the stars in $B$.  This is the [Fe/H] distribution of
the planet hosts in the synthetic SPOCS catalog.  Each iteration of this
procedure will produce a slightly different planet-host [Fe/H]
distribution, so our statistical analysis takes into account the fact
that the planet hosts in SPOCS cannot be a perfect representation of the
[Fe/H] distribution of all planet hosts in the Solar neighborhood.

\item Populate a control set $C$, with the same [Fe/H] distribution as
$B$\ and including both planetless stars and planet hosts (simulating a
collection of Pop I field stars selected independently of their planet
status):

\begin{enumerate}

\item Select a value of [Fe/H], from a uniform distribution that
stretches between the minimum and maximum values of [Fe/H] in $B$.

\item From a uniform distribution, select a probability of inclusion in
the control set $C$.

\item If the randomly selected probability of inclusion in $C$\ is less
than or equal to the value of the normalized [Fe/H]  histogram of $B$\
at the randomly selected value of [Fe/H], include in $C$\ the member of
$A$\ with the metallicity nearest this value.

\item Repeat until the control set $C$\ has the same number of stars as
the set of planet hosts $B$.
\end{enumerate}

\item Use the K-S test to calculate $k$, the probability that $B$\ and
$C$\ have the same underlying [X/Fe] distribution.

\item Use the K-S test to calculate $q$, the probability that $B$\ and
$C$\ have the same underlying [Fe/H] distribution.  Again, ${\mathcal
Q}$, the distribution underlying $q$, is unity by construction.  In
practice, $q$ will always be lower than 1, because of the impossibility
of selecting two distinct sets of stars with exactly matching [Fe/H]
distributions from a data set with a finite number of members.

\item Repeat until 100,000 Monte Carlo simulations have been performed,
then find $D_{\rm X}$\ and $D_{\rm Fe}$\ by creating histograms of the
values of $k$\ and $q$\ returned by each iteration.  After 100,000
simulations, $D_{\rm X}$\ and $D_{\rm Fe}$\ become well populated,
smooth distributions suitable for comparison.

\end{enumerate}

\section{Statistical Methods: Determining Exponent $b$\ of
Planet-Silicon Abundance Power Law}
\label{appendix2}

We used the following procedure to find $H$\ for each value of $b$:

\begin{enumerate}

\item Generate a synthetic SPOCS compilation of stellar abundances $A$,
with the same number of stars as SPOCS (1040), by sampling with
replacement.  (For details of sampling with replacement, see
\ref{appendix1}.)  Ignore any known planets: Stars in $A$\ will be
assigned synthetic planets based on [Fe/H] and [Si/Fe].

\item Select the value of $b$, the power-law exponent in equation
\ref{psieq}, that you wish to test.

\item Calculate ${\mathcal F}({\rm [Fe/H],[Si/Fe]})$, the probability
that a planet will be found around a star of a given [Fe/H] and [Si/Fe]:

\begin{enumerate}

\item Separate the stars in $A$\ into [Fe/H] bins, $i$, of width $0.1$
dex.  In each bin $i$, set ${\mathcal F} = 0$ outside the minimum and
maximum values of [Si/Fe].

\item Find the number of stars in $i$, $N_i$, and calculate the value of
the planet-metallicity correlation of equation \ref{pmceq} in each bin.

\item Separate the stars in $i$ into [Si/Fe] bins, $j$, of width $0.05$
dex.  Create a histogram of [Si/Fe] values in $i$, $h_j$.

\item Set the planet-silicon correlation such that \[ {\mathcal F}_{i,j}
= C \times 10^{b \, {\rm [Si/Fe]}_j} .\]  Find $C$ such that the {\it
average} of the nonzero values of $\mathcal F$ in $i$, is equal to the
value of the planet-metallicity correlation in $i$: \[ C_i = { 0.03
\times N_i \, 10^{2.0 \, {\rm [Fe/H]}_i} \over \sum_j h_j \, 10^{b \,
{\rm [Si/Fe]}_j} } . \]

\end{enumerate}

\item Assign planets to the stars in $A$, to form the set of planet
hosts $B$:

\begin{enumerate}

\item From a uniform distribution, find a random deviate $R$,
corresponding to the probability that the $n^{th}$ star in $A$, $A_n$,
will have a planet.

\item Find ${\mathcal F}_n$, $\mathcal F$ at the values of [Fe/H] and
[Si/Fe] corresponding to $A_n$.

\item If $R \leq {\mathcal F}_n$, $A_n$ is given a planet.  If not,
$A_n$ will be planetless.

\item Repeat for all stars in $A$.  Although the particular stars that
receive planets are different for each new set $A$, the overall fraction
of stars with planets will always be $\sim 6\%$, as in SPOCS.

\end{enumerate}

\item Find the iron-abundance distribution of the synthetic planet
hosts: Calculate the normalized histogram of [Fe/H] for all stars in
$B$.

\item Select a control set, $C$, with the same number of members and the
same [Fe/H] distribution as $B$, making no reference to planet status:

\begin{enumerate}

\item Select a value of [Fe/H], from a uniform distribution with a
spread equal to the range of [Fe/H] values present in $B$.

\item Select a probability of inclusion in $C$ from a uniform
distribution.

\item If the selected probability of inclusion in $C$ is less than or
equal to the normalized [Fe/H] histogram of $B$ at the randomly selected
value of [Fe/H], include the star in $A$ with [Fe/H] nearest this value
in the control set $C$.

\item Repeat until $C$ has the same number of members as $B$.

\end{enumerate}

\item Use the K-S test to calculate the probability, $p$, that $B$ and
$C$ have the same underlying [Si/Fe] distribution.

\item Repeat for 10,000 Monte Carlo simulations, then find $H$ by
calculating the histogram of $p$.  Compare $H$ with $D_{\rm Si}$ to
figure out whether the chosen value of $b$ produces a planet-silicon
correlation that matches what is observed in SPOCS.

\end{enumerate}

\end{document}